\documentclass[10pt,preprint]{aastex}
\usepackage{natbib}
\usepackage{multirow}
\usepackage{upgreek}
\usepackage[colorlinks=true, pdfstartview=FitV, linkcolor=blue,citecolor=blue, urlcolor=blue]{hyperref}
\usepackage{wasysym}
\usepackage{txfonts}
\usepackage{gensymb}
\usepackage{morefloats}
\usepackage{threeparttable}
\slugcomment{version \today}
\shorttitle{AGN Feedback in the Hot Halo of NGC 4649}
\shortauthors{Paggi et al.}

\begin{document}

\title{AGN Feedback in the Hot Halo of NGC 4649}

\author{Alessandro Paggi\altaffilmark{1}, Giuseppina Fabbiano\altaffilmark{1}, Dong-Woo Kim\altaffilmark{1}, Silvia Pellegrini\altaffilmark{2}, Francesca Civano\altaffilmark{3}, Jay Strader\altaffilmark{4} {and Bin Luo}\altaffilmark{5}}
\affil{\altaffilmark{1}Harvard-Smithsonian Center for Astrophysics, 60 Garden St, Cambridge, MA 02138, USA: \href{mailto:apaggi@cfa.harvard.edu}{apaggi@cfa.harvard.edu}\\
\altaffilmark{2}Department of Astronomy, University of Bologna, via Ranzani 1, 40127 Bologna, Italy\\
\altaffilmark{3}Department of Physics and Yale Center for Astronomy and Astrophysics, Yale University, P.O. Box 208121, New Haven, CT 06520-8121\\
\altaffilmark{4}Department of Physics and Astronomy, Michigan State University, East Lansing, MI 48824, USA\\
\altaffilmark{5}{Department of Astronomy \& Astrophysics, 525 Davey Lab, The Pennsylvania State University, University Park, PA 16802, USA}}

\begin{abstract}
Using the deepest available \textit{Chandra} observations of NGC 4649 we 
find strong evidences of cavities, ripples and ring like structures in the 
hot interstellar medium (ISM) that appear to be morphologically related 
with the central radio emission. {These structures show no 
significant temperature variations in correspondence with higher pressure 
regions (\(0.5\mbox{ kpc}<r<3\mbox{ kpc}\)).}
%These structures are reminiscent of the structures observed in NGC 1275 \textit{Chandra} images. In common with that source, we found no significant temperature variations in correspondence with higher pressure regions (\(0.5\mbox{ kpc}<r<3\mbox{ kpc}\)), suggesting that the observed structures may be isothermal waves whose energy is dissipated by viscosity. 
On the same spatial scale, a 
discrepancy between the mass profiles obtained from stellar dynamic and 
\textit{Chandra} data represents the telltale evidence of a significant 
non-thermal pressure component in this hot gas, which is related to the 
radio jet and lobes. On larger scale we find agreement between the mass 
profile obtained form \textit{Chandra} data and planetary nebulae and 
globular cluster dynamics. The nucleus of NGC 4649 appears to be extremely 
radiatively inefficient, with highly sub-Bondi accretion flow. Consistently 
with this finding, the jet power evaluated from the observed X-ray cavities 
implies that a small fraction of the accretion power calculated for the 
Bondi mass accretion rate emerges as kinetic energy. Comparing the jet 
power to radio and nuclear X-ray luminosity the observed cavities show 
similar behavior to those of other giant elliptical galaxies.
\end{abstract}


\section{Introduction}\label{intro}

Evidence of the interaction of Active Galactic Nuclei (AGN) with the 
surrounding hot gas in nearby galaxies and clusters has been observed as 
morphological disturbances in the X-ray halos in the form of ripples and 
cavities 
\citep[e.g.][]{2000MNRAS.318L..65F,2003MNRAS.344L..43F,2005ApJ...635..894F,2006MNRAS.366..417F}. 
The AGN-induced disturbances have also been observed 
in the hot interstellar medium (ISM) in the halos of a number of normal 
elliptical galaxies \citep[e.g.][]{2007ApJ...668..150D}, and are 
interpreted as a consequence of the thermal X-ray emitting gas being 
displaced by the AGN jets.

NGC 4649, also known as M60, is a nearby\footnote{We assume a distance to 
NGC 4649 of (\(16\mbox{ Mpc}\)). At this distance \(1''\) corresponds to 
\(77\mbox{ pc}\).} X-ray-bright giant elliptical galaxy located in a group 
at the eastern edge of the Virgo cluster. Its companion, located at \(\sim 
2.5'\) to the northwest, is the spiral galaxy NGC 4647. NGC 4649 harbors a 
faint nuclear radio source \citep{2002AJ....124..675C}. Although earlier 
\textit{Chandra} data revealed a relaxed X-ray morphology close to 
hydrostatic equilibrium - first reported by \citet{1995ApJ...452..522B} - 
there has been a debate on the presence of inhomogeneities correlated with 
the nuclear radio source. 

{Finger-like structures in the inner \(\sim 5\) kpc of the diffuse 
X-ray emission from NGC 4649 have been reported by 
\citet{2004ApJ...600..729R,2006ApJ...636..200R} in their study of \textit{Chandra} and \textit{XMM-Newton} data. These structures are compared by the authors with those predicted by hydrodynamical
simulations of cooling flows in elliptical galaxies \citep{1998MNRAS.301..343K}, that is, brighter, cooler
inflowing gas surrounded by fainter, hotter outflowing jets. However, the authors found no significant temperature variations across the observed structures.} \citet{2008MNRAS.383..923S}, using {the same \(\sim 37\mbox{ ksec}\) \textit{Chandra} observation of \citet{2004ApJ...600..729R} analysis}, found morphological 
disturbances in the X-ray emitting gas, and interpreted them as the result 
of interaction with the central AGN. Instead a subsequent analysis of 
deeper \textit{Chandra} observations by \citet{2008ApJ...683..161H} showed 
a generally undisturbed X-ray morphology, consistent with that expected 
from a hot ISM in hydrostatic equilibrium. Later, the analysis by 
\citet{2010MNRAS.404..180D} of \textit{Chandra} observations shallower that 
those of \citeauthor{2008ApJ...683..161H} - but deeper than 
\citeauthor{2008MNRAS.383..923S} - revealed (again) disturbances and 
cavities in the ISM connected with the radio emission.

Thanks to the relatively small distance and large supermassive black hole 
(SMBH) mass (\(\sim\) few \({10}^9\,M_{\astrosun}\)) of NGC 4649, 
\textit{Chandra} resolves radii close to the Bondi accretion radius, 
\(r_{acc}\approx 100-200\mbox{ pc}\), at which the gravitational binding 
energy of a gas element becomes larger than its thermal energy 
\citep{1952MNRAS.112..195B}. Thus, NGC 4649 represents an ideal case to 
investigate the following questions: what is the mass accretion rate? What 
fraction of the accretion power is in the observed nuclear luminosity, and 
what in the observed jet power? Is NGC 4649 consistent with the previously 
found correlations between the Bondi mass accretion rate \(\dot M_{B}\), 
the power associated with the observed cavities \(P_{cav}\), and radio 
luminosity \citep{2010ApJ...720.1066C,2013MNRAS.432..530R}?  Due to its low 
radio power, and its non-being a central dominant galaxy, NGC 4649 is also 
an ideal case to investigate whether these correlations, mostly found for 
radio-bright central dominant galaxies in groups or clusters, work equally 
well in more ``normal'' elliptical galaxies.

In this paper we revisit the properties of the hot ISM of NGC 4649, using 
much deeper \textit{Chandra} data with respect to previous studies and 
updated atomic databases. We find strong evidence of cavities, ripples and 
ring like structures that appear to be morphologically related with the 
central radio emission. In addition, we find that the hot halo is subject 
to an additional non-thermal pressure term as already reported in previous 
studies
\citep[e.g.][]{2009ApJ...705.1672B,2010MNRAS.409.1362D,2011MNRAS.415.1244D,2013MNRAS.430.1516H}. 
We show that the non-thermal pressure is found on the 
same scale spatial as the disturbances of the halo and it is spatially 
correlated with the hot gas pressure and the minimum pressure derived from 
the radio data. The nucleus of NGC 4649 appears to be extremely radiatively 
inefficient, with highly sub-Bondi accretion flow, releasing a very small 
fraction of the accretion power in form of kinetic energy in the 
surrounding halo.

The paper is organized as follows: Section \ref{sec:data} describes the 
data sets used in this work, the reduction procedures, and the image and 
spectral analysis. In Section \ref{sec:discussion} we discuss our results, 
and Section \ref{sec:summary} is dedicated to our conclusions.

\section{Data Reduction and Analysis}\label{sec:data}

NGC 4649 has been observed by \textit{Chandra} with the ACIS detector 
\citep{1997AAS...190.3404G} six times, between April 2000 and August 2011. 
Level 2 event data were retrieved from the \textit{Chandra} Data 
Archive\footnote{\href{http://cda.harvard.edu/chaser}{http://cda.harvard.edu
/chaser}} and reduced with the CIAO (\citealt{2006SPIE.6270E..60F}) 4.5 
software and the \textit{Chandra} Calibration Data Base (\textsc{caldb}) 
ver. 4.5.7, adopting standard procedures. After excluding time intervals of 
background flares exceeding \(3\sigma\) with the \textsc{lc\_sigma\_clip} 
task, we obtained the low-background exposures listed in Table 
\ref{table:obs}, for a total exposure of \(\sim 280\) ks.

For each data set we generated a full resolution image in the 
\(0.3-8.0\mbox{ keV}\) energy band. To take advantage of the longer 
exposure time and identify fainter signatures we also produced a merged 
image of the six observations; to this end we used the \textsc{wavdetect} 
task to identify point sources in each observation with a {2} 
sequence of wavelet scales (i.e., 1, 2, 4, 8, 16 and 32 pixels) and a 
false-positive probability threshold of \({10}^{-6}\). Then we used the 
\textsc{reproject\_aspect} task to modify the aspect solution minimizing 
position differences between the sources found, and finally merged the 
images with the \textsc{merge\_all} script.

\subsection{Image of halo}\label{sec:image}

We examined the co-added image for evidence of morphological disturbances 
that may be connected with the nuclear radio source. 
{To this end we then ran \textsc{wavdetect} on the merged image to detect fainter point sources. For each of these 
appropriate elliptical regions were generated with the \textsc{roi} CIAO 
task, both for the source and for the nearby background. We then} 
processed the merged image with the \textsc{dmfilth} task to remove the 
detected point sources, replacing the counts by sampling the Poisson 
distribution of the pixel values in the concentric background region. The 
resulting data image is therefore expected to show the morphology of the 
ISM diffuse emission, although it will still include point sources below 
the detection threshold.

\begin{figure}
\centering
\includegraphics[scale=0.7]{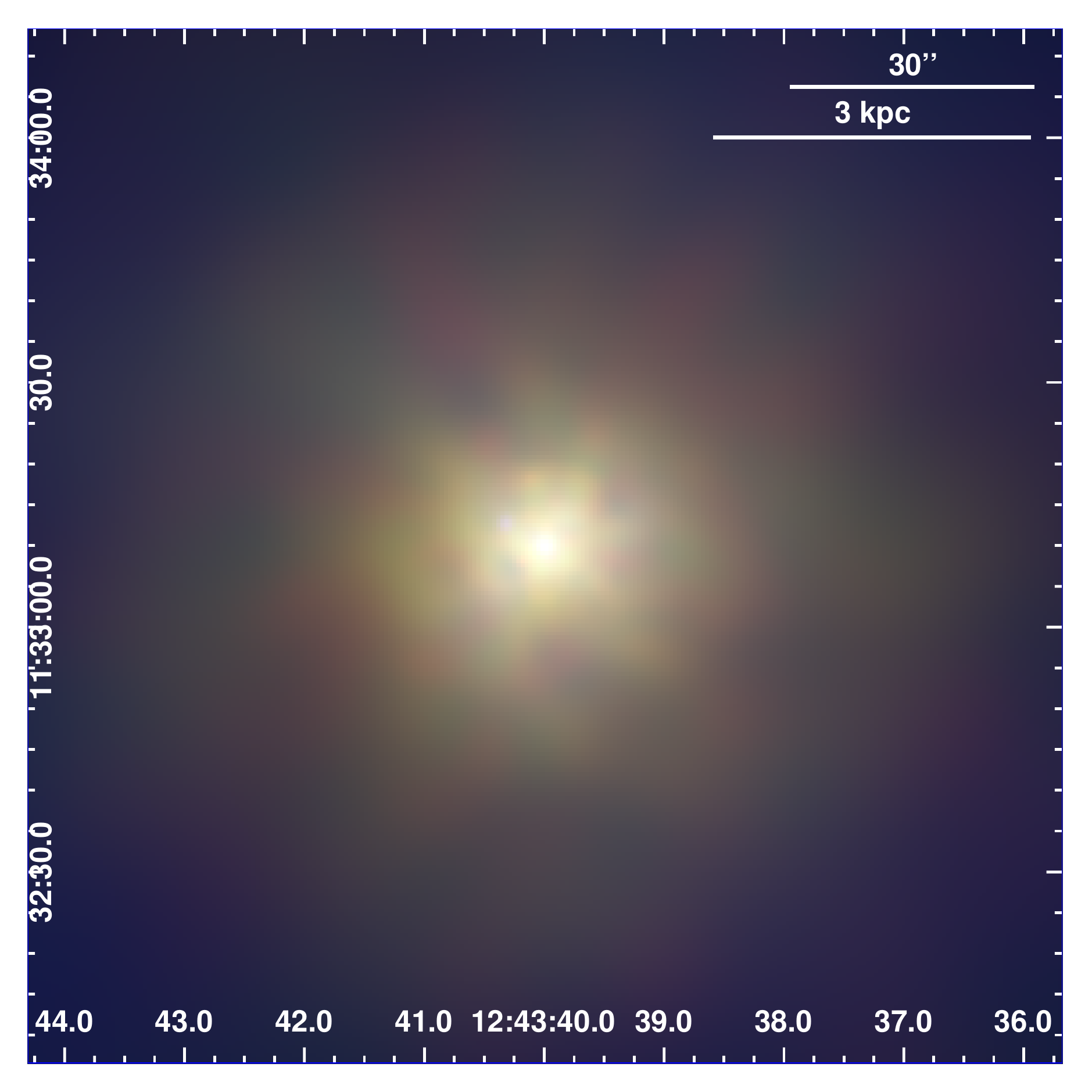}
\caption{Merged three-color \textit{Chandra} ACIS-S image of the inner 
\(\sim 5\mbox{ kpc}\) region NGC 4649. The three energy bands are shown 
with different colors, that is, \(0.5-1.0\mbox{ keV}\) (red), 
\(1.0-2.0\mbox{ keV}\) (green), and \(2.0-8.0\mbox{ keV}\) (blue). Each 
band image is adaptively smoothed using the \textsc{csmooth} tool. Besides 
some hard, non detected (or not completely removed) point sources, the 
image shows {hints} of structures and cavities in the soft 
emission.}\label{fig:color}
\end{figure}

We produced separate images in different energy bands (\(0.5-1.0\mbox{ 
keV}\), \(1.0-2.0\mbox{ keV}\) and \(2.0-8.0\mbox{ keV}\)) in order to 
obtain a three-color image that may pinpoint evidences of structures in the 
ISM. We then adaptively-smoothed each band image using the \textsc{csmooth} 
tool \citep{2006MNRAS.368...65E}, with minimum and maximum significance S/N 
levels of \(4.5\) and \(5.5\), respectively to enhance fainter, extended 
features of the diffuse emission. The merged three-color image (Figure 
\ref{fig:color}) suggests hints of structures and cavities in the soft 
emission in the inner \(\sim 3\mbox{ kpc}\) region.

\begin{figure}
\centering
\includegraphics[scale=0.8]{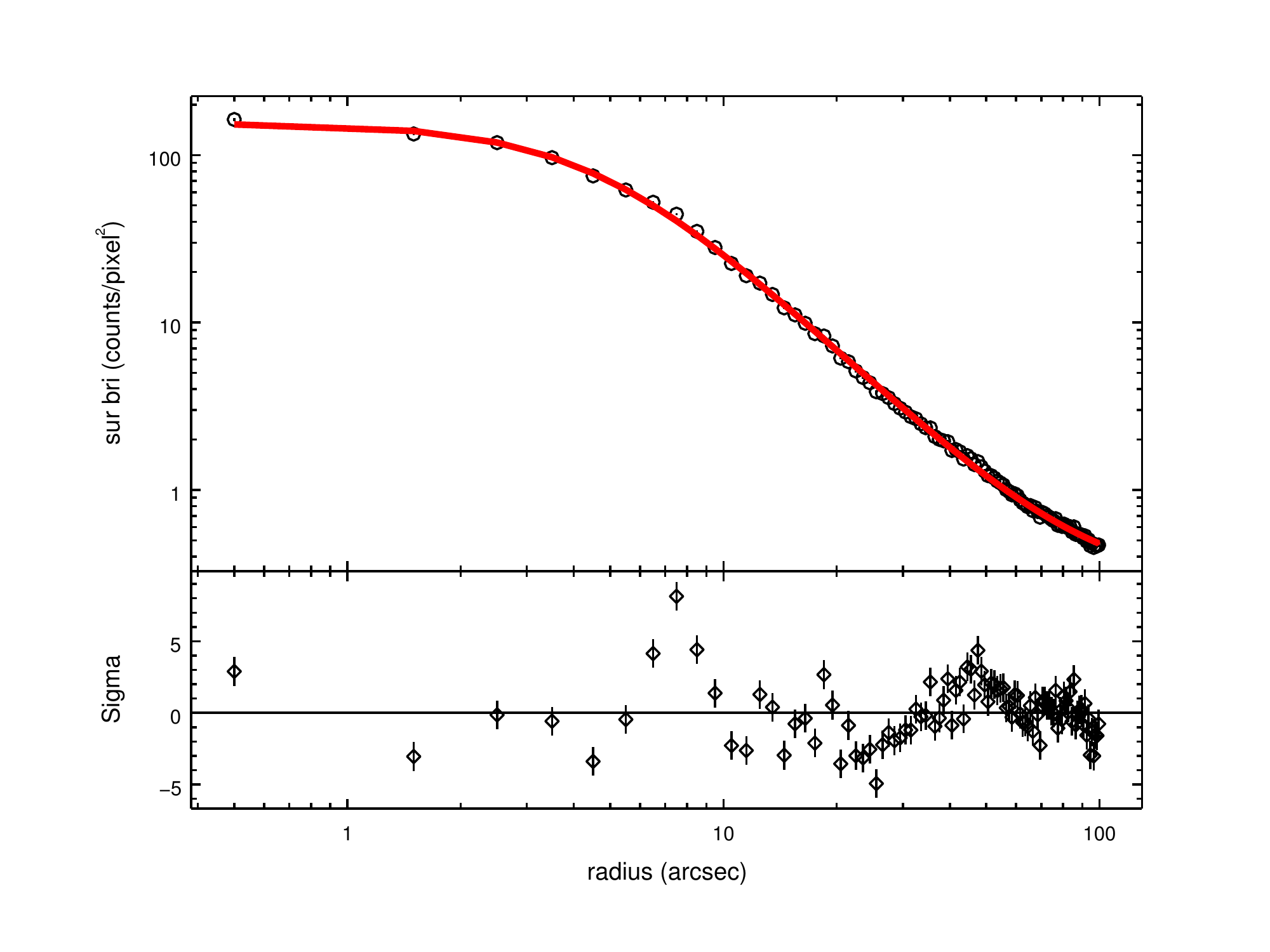}
\caption{Radial profile of NGC 4649 \(0.3-8\mbox{ keV}\) surface 
brightness, with superimposed the best fit \(\beta\) model fit. As 
discussed in the main text, the fit is very poor (with \(\chi^2\sim 4\)) 
and shows significant residuals at \(\sim 8\)'' and \(\sim 
40\)''.}\label{rprofile}
\end{figure}

To gain a more quantitative insight on morphological disturbances, we 
performed a one-dimensional fit of the surface brightness profile - 
evaluated in concentric annuli in the \(0.3-8\mbox{ keV}\) source-free 
merged image - with a standard \(\beta\) model using \textsc{Sherpa}. The 
fit was performed in the inner 100'' (\(\sim 8\mbox{ kpc}\)) in order to 
avoid contamination from the emission of the companion galaxy NGC 4647. The 
results (Figure \ref{rprofile}) show that, although the \(\beta\) model 
reproduces the large-scale behavior of the surface brightness profile, the 
fit is poor, with a reduced \(\chi^2\sim 4\). Adding a second \(\beta\) 
model did not increase the goodness of the fit, while yielding unphysical 
model parameters. Most of the \(\chi^2\) contribution comes from well 
defined radii at \(\sim 8\)'' and \(\sim 40\)'' (corresponding to \(\sim 
0.5\mbox{ kpc}\)  and \( 3\mbox{ kpc}\), respectively).

\begin{figure}
\centering
\includegraphics[scale=0.40]{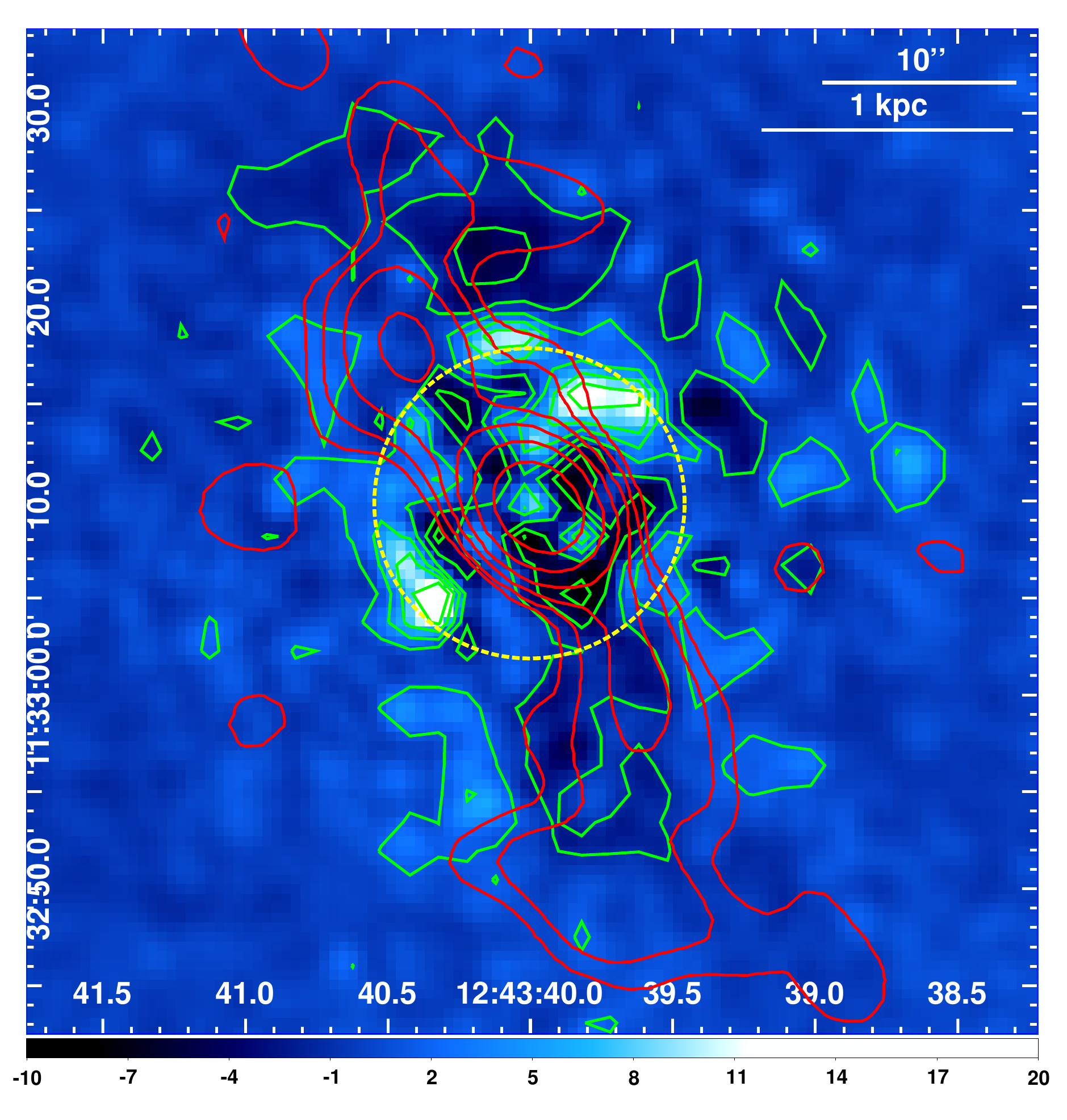}
\includegraphics[scale=0.40]{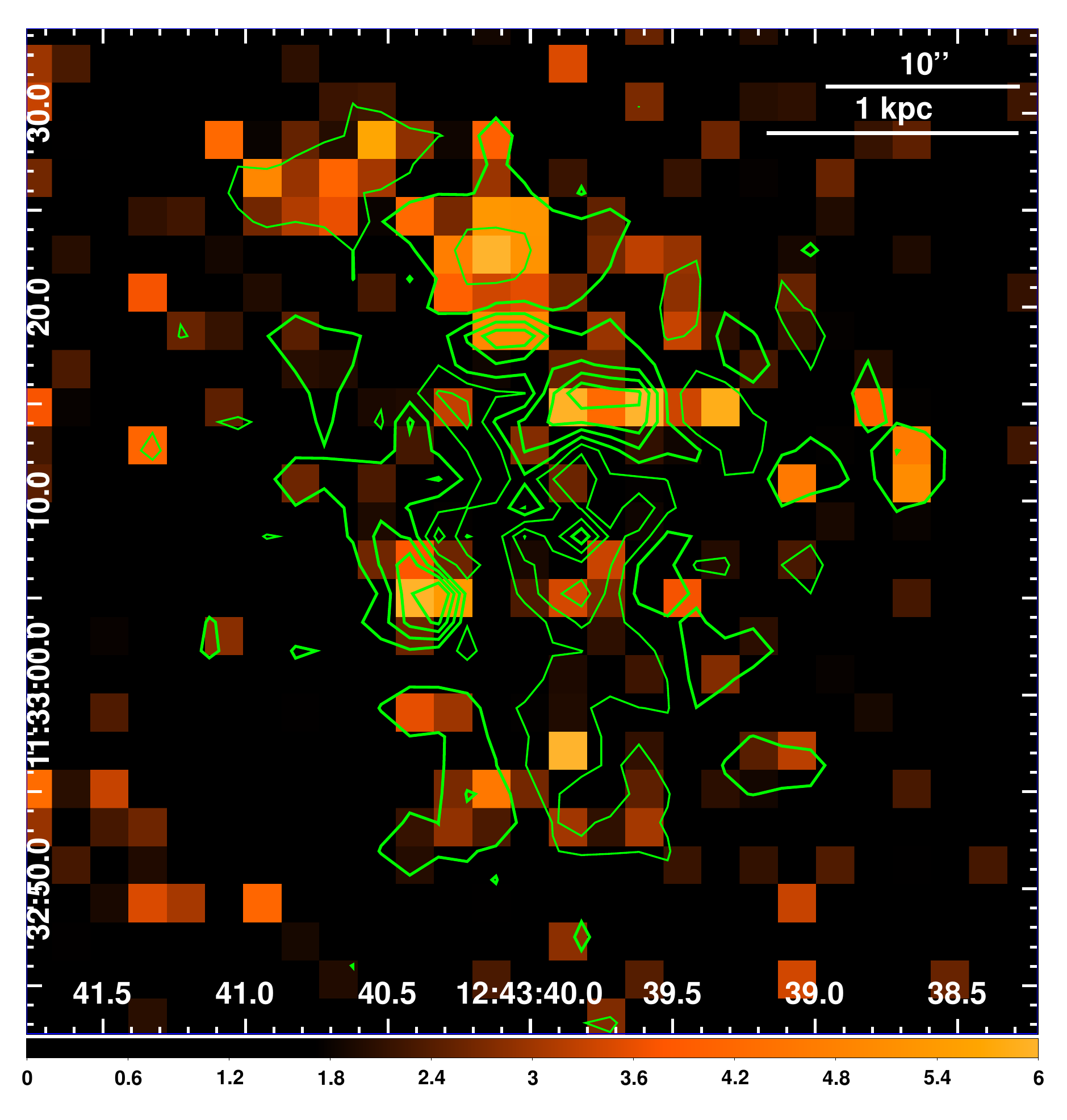}
\caption{(left panel) Distribution of residuals relative to an elliptical 
\(\beta\) model of the inner \(\sim 1 \mbox{ kpc}\) region with a 3X3 FWHM 
gaussian smoothing, with contours shown in green and VLA emission contours 
are shown in red. The dashed yellow circle indicates the \(\sim 8''\) 
radius where we see significant residuals in the one-dimensional surface 
brightness fit (see Figure \ref{rprofile}). (right panel) Residual S/N map, 
evaluated as the ratio between the residuals and the X-ray counts error, 
binned to a pixel size 4 times the size of the native ACIS-S pixel. Green 
contours are the same as in left panel.}\label{fig:sign3}
\end{figure}

To investigate the origin of these residuals, we performed a 2-D fit of the 
full-band source-free merged image with a two-dimensional \(\beta\) model. 
Again, adding a second \(\beta\) model in our fit did not increase the fit 
goodness significantly. The residual distribution in the inner \(\sim 
1\mbox{ kpc}\) region is presented in the left panel of Figure 
\ref{fig:sign3}, where we applied a 3 pixel FWHM Gaussian smoothing to 
highlight the presence of structures. In the same panel we superimpose to 
the residual contours (shown in green) the \(1.4\mbox{ GHz}\) VLA emission 
contours \citep[][shown in red]{1986MNRAS.220..363S}. We notice a striking 
correlation between the outer radio lobes and the regions of negative 
residuals; regions of positive residuals appear to lie on both sides of the 
radio emission and are responsible for the features at a galactocentric 
radius of \(\sim 8''\) seen in Figure \ref{rprofile}. On a larger scale 
(\(\sim 40''\)), the residual map shows a ring-like structure - recalling 
those observed in NGC 1275 \citep[e.g.,][]{2006MNRAS.366..417F} - as shown 
in \ref{fig:sign10}, where on the left panel we present the inner \(\sim 
5\mbox{ kpc}\) region, where we applied a 10 pixel FWHM Gaussian smoothing. 
Radio emission contours are superimposed as in Figure \ref{fig:sign3}.
The significance of these structures is shown in the right panels of 
Figures \ref{fig:sign3} and \ref{fig:sign10}. Here we show the residual S/N 
map, evaluated as the ratio between the residuals and the X-ray counts 
error, binned to a pixel size 4 and 14 times the size of the native ACIS-S 
pixel, respectively. At this pixel size the residual S/N map closely 
follows that of the residual map with the gaussian smoothing, as shown by 
the superimposed contours of the residual map (shown in green). In 
particular, we notice that the S/N of the binned pixel in the observed 
structures is of the order of 3 or higher, pointing to an higher 
significance of the structures as a whole.

{In order to evaluate the reliability of the observed structures and 
their dependence on the particular model used, we performed the 2-D fitting 
procedure described above excluding data between 5'' and 70'' (so avoiding 
the structures shown in Figure \ref{fig:sign10}) and then subtracted the 
best-fit model so obtained from the complete dataset. In addition, we 
repeated the same procedure in the four 90 degrees quadrants to investigate 
possible azimuth anisotropies that could yield the observed structures. All 
the residual maps so obtained are very similar and show the same structures 
presented in Figures \ref{fig:sign3} and \ref{fig:sign10}. To investigate 
the effect of the point-source removal procedure described above on the 
observed residual structure, we also used the source catalog produced by 
\citet{2013ApJS..204...14L} (which used a finer \(\sqrt{2}\) sequence of wavelet 
scales), obtaining again very similar results. We therefore conclude that 
the observed structures are real and do not strongly depend on the details 
of our analysis.}

\subsection{Spatially resolved spectral analysis}\label{sec:spectra}

To estimate the properties of the hot gas in the regions of enhanced and 
suppressed X-ray emission, we performed spectral analysis. Figure 
\ref{fig:outer_regions} shows the region used for spectral extraction with 
the CIAO \textsc{specextract} task. For each extraction region, background 
spectra were extracted in the same region from appropriate ``blank-sky" 
fields and normalized equating the \(9-12\mbox{ keV}\) count rates of the 
observed and background data, since essentially all those hard X-rays are 
due to particle background \citep{2006ApJ...645...95H}. Additionally, we 
extracted spectra in a series of concentric, contiguous annuli, with widths 
chosen so as to contain approximately the same number of 
background-subtracted photons (\(\sim 12000\)). We placed a lower limit of 
2.5'' on the annulus width to ensure that the instrumental spatial 
resolution does not lead to strong mixing between the spectra in adjacent 
annuli. To account for projection effects, we used the \textsc{projct} model 
implemented in \textsc{Xspec} \citep[ver. 12.8.0,][]{1996ASPC..101...17A}. 
We excluded data in the vicinity of any detected point source, as well as in 
the central part of the interloper galaxy NGC 4647 to prevent possible 
contamination. We extracted data individually for each observation and 
combined the source and background spectra, generating spectral response 
matrices weighted by the count distribution within the aperture (as 
appropriate for extended sources). To make use of the \(\chi^2\) fit 
statistic we binned the spectra to obtain a minimum of \(20\) counts per bin 
using the \textsc{specextract} task; in the following, errors correspond to 
the \(1\)-\(\sigma\) confidence level for one interesting parameter 
(\(\Delta\chi^2 = 1\)). In all the spectral fits we included photo-electric 
absorption by the Galactic column density along the line of sight \(N_H = 
2.04\times {10}^{20}\mbox{ cm}^{-2}\) \citep{2005A&A...440..775K}.

Spectral fitting were performed in the \(0.3-8\mbox{ keV}\) energy range 
using a model comprising a \textsc{vapec} thermal component, plus a thermal 
bremsstrahlung component to account for undetected point sources 
\citep{2003ApJ...587..356I}. We adopted solar abundances from 
\citet{1989GeCoA..53..197A}, and we allowed the global ratios of O, Ne, Mg 
and Si with respect to Fe to fit freely
\citep[see e.g.][]{2012ApJ...751...38K}, and fixed the remaining ratios at 
the solar value. In addition, as NGC 4649 is within the Virgo cluster we 
included an additional hot gas component, with kT fixed at 2.5 keV 
\citep[e.g.][]{2002ApJ...572..160G} to account for possible interloper 
cluster emission. The results are presented in Table 
\ref{table:fit_results}. The details of the element abundances obtained in 
the fits will be discussed in a forthcoming paper \citep{paggi2014}. Here we 
concentrate on the physical parameters (temperature and gas density). In 
Figure \ref{fig:spectral_results} we show gas temperature and projected 
pressure \(P_{{proj}}=kT\,A^{1/2}\) (where \(A\) is the \textsc{vapec} 
component normalization per arcsec square) in the extraction regions (shown 
in the corresponding colors) in comparison with the average values obtained 
in the annulus at the same radius (shown in grey). There is no significant 
temperature variation between overdense and underdense regions. Instead, the 
projected pressure is higher in brighter regions.

Figure \ref{fig:profiles} shows temperature and gas density profiles out to 
a radius \(\sim 20\mbox{ kpc}\) in comparison with that obtained with 
similar techniques by \citet{2008ApJ...683..161H} using shallower 
\textit{Chandra} data. The temperature we derive is systematically higher 
than that of \citet{2008ApJ...683..161H}, while our gas density is 
systematically lower. This effect is manly due to the use of the updated 
ATOMDB 2.0.2 with respect to the 1.3.2 version available before 2010 
\citep[see][]{2012ApJ...757..121L}. In particular with respect to this 
previous analysis, while the gas density profiles have similar shapes, we 
see that our temperature profile, rather than a decrease up to \(\sim 
1.5\mbox{ kpc}\) followed by a monotonic increase up to the outer regions, 
is characterized by a decrease up to \(\sim 0.5\mbox{ kpc}\) followed by a 
``plateau" of almost constant temperature up to \(\sim 3\mbox{ kpc}\) (the 
same scales of the structures shown in Figure \ref{fig:sign10}). The two 
profiles are properly fitted by a double broken power-law model (shown in 
Figure \ref{fig:profiles} with the dashed lines), with breaks at \(\sim 
7.6''(0.6\mbox{ kpc})\) and \(\sim 42.2''(3.2\mbox{ kpc})\).

\section{Discussion}\label{sec:discussion}

In the local universe, most central SMBHs in elliptical galaxies are 
radiatively quiescent 
\citep{2005ApJ...624..155P,2010ApJ...717..640P,2006ApJ...640..126S,2008ARA&A..46..475H,2010ApJ...714...25G}, 
and often seen in a ``radio-mode'', where they are able to drive jets 
\citep[e.g.,][]{2002ApJ...564..120H,2007MNRAS.381..589M}. Mechanical 
feedback of these jets has a significant impact on the surrounding ISM, with
implications for galaxy and SMBH coevolution 
\citep[e.g.][]{2005Natur.433..604D,2006MNRAS.365...11C,2007MNRAS.380..877S,2010ApJ...722..642O,2013SSRv..tmp...85K}, 
and regulation of the hot gas cooling \citep{2010MNRAS.404..180D}. Many 
aspects of this feedback process, though, are not well understood, including 
the nature of the material feeding the SMBHs, the details of how accretion 
originates a mechanical energy output, in the form of a wind or a jet, and 
then how this couples to the surrounding ISM heating it. \textit{Chandra} 
observations of hot gas rich elliptical galaxies often show direct evidence 
of mechanical feedback in the form of radio jets and inflated bubbles 
displacing the hot plasma \citep[e.g.][]{2007ARA&A..45..117M}. From a study 
of nine X-ray luminous ellipticals harboring cavities, sufficiently nearby to
measure the hot gas density and temperature reasonably close to the SMBH, a 
tight coupling was found between the Bondi mass accretion rate  ($\dot 
M_{B}$) and the jet power estimated from the energy associated with the 
observed cavities $P_{cav}$ 
(\citealt{2006MNRAS.372...21A,2007MNRAS.381..589M}, see also 
\citealt{2013MNRAS.432..530R}). $P_{cav}$ correlates also with the 
(extended) radio luminosity, albeit in relationships showing a large scatter 
\citep{2008ApJ...686..859B,2010ApJ...720.1066C,2011ApJ...735...11O}.

Thanks to the long cumulative \textit{Chandra} pointing on NGC 4649 we have 
determined the hot gas density and temperature in a wide range of radii, 
discovering cavities and a ring-like higher-pressure ripple. We also 
determine these parameters in the nuclear region with unprecedented accuracy.

\subsection{Temperature profile}
The residual map of the X-ray emission with respect to a standard \(\beta\) 
model (Figure \ref{fig:sign10}) shows significant cavities and bright 
spots on sub-kpc scale, as well as more extended ring like structures at 
scales \(\sim 3\mbox{ kpc}\). These previously unreported features appear 
to be morphologically related with the weak nuclear radio source 
\citep{2002AJ....124..675C,2008MNRAS.383..923S}. We find no significant 
temperature variations between under-dense and over-dense regions (see 
Figure \ref{fig:spectral_results}). Moreover, looking at the average 
temperature profile presented in Figure \ref{fig:profiles} (left panel), we 
see an almost constant temperature \(\sim 0.85\mbox{ kT}\) in the \( 0.5 
\div 3\mbox{ kpc}\) range. This resembles the case of the ripples and 
cavities around NGC 1275 in the Perseus cluster, discovered by 
\citet{2006MNRAS.366..417F} with \textit{Chandra}, which also shows no sign 
of increased temperature in higher pressure regions. 
\citeauthor{2006MNRAS.366..417F} concluded that the NGC 1275 structures may 
be isothermal waves whose energy is dissipated by viscosity, with thermal 
conduction and sound waves effectively distributing the energy from the 
radio source.

The radial kT profile of NGC 4649 shows a sudden increase up to \(kT\sim 
1.25\mbox{ keV}\) in the inner \(0.5\mbox{ kpc}\). \textit{Chandra} 
observations revealed central temperatures higher than the surroundings in 
other early type galaxies with low or very low power radio sources: NGC 4594 
\citep{2003ApJ...597..175P}, NGC 4552 \citep{2006ApJ...648..947M}, NGC 3115 
\citep{2011ApJ...736L..23W} and NGC 4278 \citep{2012ApJ...758...94P}. This 
hotter core has been interpreted as a result of heating from  gravitational 
effects due to the central SMBH, a recent AGN outburst, or interaction with 
confined nuclear jets. The clear evidence for AGN feedback on the large 
scale in NGC 4649 (i.e., the series of cavities, likely more numerous than 
the N and S ones that will be investigated in detail in Sect. 
\ref{sec:jet_power}) provides support to the idea of repeated AGN heating 
occurring at the center, possibly producing the temperature peak, for 
example through shock activity. Note that the radiative cooling time in the 
innermost radial bin is just \(t_{cool}\sim 2.9\times{10}^7\) yr \citep[{using} the 
cooling curve from][]{2005MNRAS.358..168S}, then if such cooling is to be 
offset a heating mechanism is required. {Following the calculation by \citet{2006MNRAS.366..417F}, we can interpret the observed structures as isothermal waves and show that they can provide an effective way to heat the ISM. In fact, the thermal pressure deviations from the average value (shown in Figure \ref{fig:spectral_results}) are about 5-10\% (comparable with those observed in NGC 1275, although with less significance). Then, such waves can balance radiative cooling if the cooling time is \(\sim 20\) times the crossing time evaluated from the local sound speed \(\sim 500\mbox{ km}\mbox{s}^{-1}\) \citep{2004ApJ...607..800B}. This condition is met, since the cross time for the inner \(\sim 5\mbox{ kpc}\) region is \(t_{cross}\sim 1\times{10}^7\mbox{ yr}\), while the cooling time at this radius turns out to be \(t_{cool}\sim 5\times{10}^8\mbox{ yr}\). So, even if constraints on isothermal waves are not as tight as in the case of NGC 1275 due to larger uncertainties, it is possible for them to be the source of the ISM disturbances we observe in NGC 4649.
%Also for the innermost \(200\mbox{ pc}\) bin we have \(t_{cross}\sim 5\times{10}^5\mbox{ yr}\) and \(t_{cool}\sim 3\times{10}^7\mbox{ yr}\)
}

\subsection{Non-thermal pressure component}\label{sec:ntpress}

To further investigate the origin of the observed ISM disturbances, we 
performed an analysis similar to that proposed by \citet{2013MNRAS.430.1516H}, confronting the mass profiles obtained from the 
hot gas profiles presented in Figure \ref{fig:profiles} through the 
hydrostatic equilibrium equation and the ones obtained from independent 
diagnostics such as stellar kinematics, globular clusters (GCs) and 
planetary nebulae (PNe) velocities 
\citep{2010ApJ...711..484S,2012ApJ...748....2D}. The profiles are presented 
in Figure \ref{fig:masses} (left panel), the hot gas-derived profile is in 
light blue and that from stellar and GC kinematics in red 
\citep{2010ApJ...711..484S}, with the strips width representing the relative 
uncertainties. In the outer regions the mass profiles from 
\citet{2012ApJ...748....2D} are plotted in green (GCs) and dark blue (PNe).

At larger radii we notice a significant discrepancy between the mass profile 
derived by \citet{2010ApJ...711..484S} using stellar 
\citep{2003ApJ...583...92G} and GC \citep{2008ApJ...674..869H} kinematics, 
and the mass profiles obtained by \citet{2012ApJ...748....2D} using PNe 
\citep{2011ApJ...736...65T} and GC (same dataset as 
\citeauthor{2010ApJ...711..484S}) kinematics assuming spherical symmetry. We 
note that \citeauthor{2012ApJ...748....2D} report that PNe are more 
centrally concentrated and are on more radial orbits with respect to the total GC population, while the red GC subpopulation is similar to that of both the stars and PNe.
These discrepancies have been discussed by \citet{2011ApJ...736...65T}, 
\citet{2011MNRAS.415.1244D} and \citet{2013MNRAS.436.1322C}, who suggested 
that GC an PNe may represent dynamically distinct systems. 
In addition, the 
analysis of stellar kinematics observation from the SLUGSS survey by 
\citet{2013arXiv1310.2607A} shows strong evidence for NGC 4649 of a fast 
rotating embedded disk structure, pointing to this galaxy as a candidate for 
a major merger remnant. Finally, a recent analysis by \citet{dabrusco2013} 
revealed significant anisotropies in the two-dimensional distribution of GCs 
in NGC 4649, in particular in red and blue GC populations, that may affect 
these mass profiles. 
However, we note that the GC anisotropies revealed by 
\citet{dabrusco2013} are on a bigger scale than the ISM disturbances studied 
here. 

In the inner radii, the mass profile from X-ray emitting hot gas and that 
from stellar and GC kinematics differ significantly between 0.5 and 3 kpc, 
again the same scales of the structures shown in Figure \ref{fig:sign10}. In 
particular, the blue profile has been obtained using the equation
\begin{equation}\label{equ:he}
M_{HE}(<R) = -R\,\frac{kT_g(R)}{G\mu m_p}\left({\frac{d \log{\rho_g}}{d 
\log{R}}+\frac{d \log{T_g}}{d \log{R}}}\right)\, ,
\end{equation}
where \(R\) is the radius, \(T_g\) is the gas temperature, \(\rho_g\) is the 
gas densities, \(m_p\) the proton mass and \(\mu\approx 0.62\) is the 
average molecular weight factor. A non-thermal pressure term can, however, 
be added to Eq. \ref{equ:he}, that is
\begin{equation}\label{equ:he_ntp}
M(<R) = M_{HE}(<R) -\frac{R^2}{G\,\rho_g(R)}\frac{d P_{NT}}{d R}\, .
\end{equation}
Assuming that the stellar kinematics profile accounts for the total mass 
expressed by Eq. \ref{equ:he_ntp}, we can estimate the non-thermal pressure 
term writing
\begin{equation}\label{equ:ntp}
\frac{d P_{NT}}{d R}=-\frac{G\,\rho_g(R)}{R^2}
\left[{M(<R)-M_{HE}(<R)}\right]\,,
\end{equation}
and \(P_{NT}\) can be obtained by numerically integrating Eq. \ref{equ:ntp}. 
In Figure \ref{fig:masses} (right panel) we show the profile of the ratio 
between the non-thermal pressure {and} the average gas pressure obtained in the annular fit, { which is consistent with that obtained by \citet{2013MNRAS.430.1516H}}. For comparison, in the same panel, 
we plot the ratio of gas pressure (i.e., excess and deficit in Figure \ref{fig:spectral_results}b) in the same regions
shown in Figure \ref{fig:outer_regions}. 
Adopting the radio lobe parameters reported by \citet{2008MNRAS.383..923S} 
we also computed the minimum radio pressure assuming energy equipartition 
between particles and magnetic field in the jet 
\citep[see, e. g.,][]{2004ApJ...612..729H}, and plotted it for N and S lobes 
in Figure \ref{fig:masses} (right panel) as black squares. Although there 
are large uncertainties in these estimates, the radio jet pressure appears 
comparable with the thermal gas pressure at similar radii.

We note that the non thermal pressure can account on average for {\(\sim 
30\%\)} of the gas pressure in the regions of significant ISM disturbance, 
and that there is a striking correlation between peaks and dips in the two 
pressure trends. A simple cross-correlation test between the two profiles 
yields that without any lag the correlation is significant at the \(\sim 
99\%\), indicating a correlation between the two profiles. This links the 
non-thermal pressure to the nuclear radio source and its jets. 
\citet{2013MNRAS.430.1516H} advanced several possibilities for the 
non-thermal pressure component observed in NGC 4649, including gas rotation, 
random turbulence, magnetic field and cosmic ray pressure. Our analysis 
strongly correlates the non-thermal pressure with the radio jet structure, 
and therefore we conclude that cosmic ray injection into the ISM from the 
weak radio jets is the most likely origin of this pressure component.

\subsection{Black Hole Mass determination}

To investigate the effects of the non-thermal pressure on the SMBH mass 
estimate we fitted the mass profile derived with Eq. \ref{equ:he} with the 
four standard galaxy mass components, that is, the gas component shown in 
Figure \ref{fig:profiles}, the dark matter described by a standard NFW 
profile \(\rho_{DM}=\rho_0/ \left[{{\frac{R}{R_S} 
{\left({1+\frac{R}{R_S}}\right)}^2}}\right]\), the stellar mass obtained 
from the observed V-band luminosity \citep{2009ApJS..182..216K} (with a free 
to vary \(M/L_V\)), and the central SMBH. The result of the fit procedure is 
presented in Figure \ref{fig:mass_fit}. The best fit parameters are 
\(\rho_0=9.6\,M_{\astrosun}\mbox{ kpc}^{-3}\), \(R_S=11.1\mbox{ kpc}\) and 
\(M/L_V=4.5\). These yield a SMBH mass estimate \(M_{BH}= (5.7\pm 0.7)\times 
{10}^9 M_{\astrosun}\), compatible with - but nominally larger than - the 
SMBH mass estimate of \((4.5\pm 1.0)\times {10}^9 M_{\astrosun}\) by 
\citet{2010ApJ...711..484S}. Repeating the fitting procedure using the dark 
matter logarithmic profile proposed by \citet{2010ApJ...711..484S}, we 
obtained a similar value of \(M/L_V=3.9\) but an even larger \(M_{BH}= 
(6.1\pm 1.1)\times {10}^9 M_{\astrosun}\). Besides the details of the DM 
halo (more relevant in the outer regions), comparing with the mass profile 
form \citep{2010ApJ...711..484S} we note the main effect of the non-thermal 
pressure (significant in the \(0.5\div 3\mbox{ kpc}\) range) on this fitting 
procedure is to yield higher values of \(M/L_V\), that is, a higher 
contribution of the stellar component to the total mass, and as a 
consequence a lower SMBH mass. However, since the central SMBH mass is 
mainly driven by the mass profile in innermost radii, its estimate does not 
appear to be strongly affected by the presence of the non-thermal pressure 
component. We note, however, that our SMBH mass estimate is higher than the 
\((3.4\pm 1.0)\times {10}^9 M_{\astrosun}\) value obtained by 
\citet{2008ApJ...683..161H} using shallower \textit{Chandra} X-ray data. 
This discrepancy is explained with the higher gas temperature we observe 
with respect to \citeauthor{2008ApJ...683..161H} (see Figure 
\ref{fig:profiles} and Eq. \ref{equ:he}).

\subsection{Nuclear luminosity and Bondi accretion}\label{sec:bondi_power}

To evaluate the nuclear luminosity we extracted the \textit{Chandra}-ACIS 
spectrum of the inner \(R=2''\), adding to the model described in Sect. 
\ref{sec:spectra} a power-law component with photon index \(\Gamma\) fixed 
to 2 in order to model the AGN emission 
\citep[e.g.,][]{2010ApJ...714...25G}. The nuclear luminosity within is 
therefore \(L_{0.5-8\,{\rm keV}} = 7.2_{-1.4}^{+1.3}\times{10}^{38}\mbox{ 
erg} \mbox{ s}^{-1}\), \(L_{0.3-10\,{\rm keV}} = 9.1_{-1.7}^{+1.6} 
\times{10}^{38}\mbox{ erg} \mbox{ s}^{-1}\), and \(L_{2-10\,{\rm keV}} = 
4.2_{-0.8}^{+0.7}\times{10}^{38}\mbox{ erg} \mbox{ s}^{-1}\). Given the mass 
of the SMBH, this X-ray nuclear emission corresponds to a very sub-Eddington 
bolometric emission. In fact, for a large sample of nearby low luminosity 
AGNs, the median bolometric correction is $L_{bol}/L_{2-10\, {\rm 
keV}}\approx 15.8$ \citep{2008ARA&A..46..475H}; \citet{2007MNRAS.381..589M} 
adopt $L_{bol}/L_{2-10\, {\rm keV}}=5$ for their study of 15 low luminosity 
AGNs; \citet{2007MNRAS.381.1235V} find $L_{bol}/L_{2-10\, {\rm keV}}\approx 
10$ for low-luminosity sources. The latter choice gives $L_{bol,nuc}= 
4.2\times 10^{39}$ erg s$^{-1}$, and then $L_{bol,nuc}/L_{Edd}=7.2\times 
10^{-9}$. From an estimate of the mass accretion rate $\dot M$, we can next
discuss the accretion modalities.

The simplest assumption for gas accretion is that it is steady, spherically 
symmetric, with negligible angular momentum, as in the theory developed for 
gas accreting onto a point mass \citep{1952MNRAS.112..195B}. The accretion 
rate $\dot M_B$ then comes from the gas density, temperature, and the SMBH 
mass, for which an accurate estimate is available for NGC 4649 (\(4.5\pm 1.0 
\times 10^9 M_{\odot}\), \citealt{2010ApJ...711..484S}).  Ideally, one 
should insert in the $\dot M_B$ calculation the density and temperature at
$r_{acc}=2GM_{BH}/c_s(\infty)^2$, where the dynamics of the gas start to be 
dominated by the potential of the SMBH (\(c_s(\infty)\) is the ambient sound 
speed; \citealt{2002apa..book.....F}). In practice, one uses fiducial 
temperature and density for the circumnuclear region, determined as close as 
possible to the SMBH. In our case, the gas properties are derived reasonably 
close to the black hole, thus giving a $\dot M_B$ estimate far better than 
usual \citep[e.g.][]{2005ApJ...624..155P,2013MNRAS.432..530R}. Inserting in 
$c_s(\infty)$ the temperature ($kT=1.25_{-0.01}^{+0.03}$ keV) at the 
innermost bin, that extends out to 200 pc, gives $r_{acc}=121$ pc (for the 
polytropic index $\gamma=5/3$ of the adiabatic case), or $r_{acc}=200$ pc 
($\gamma=1$, isothermal case).  For the gas density of the innermost bin,
$\rho_{gas}=4.54_{-0.12}^{+0.12}\times 10^{-25}$ g cm$^{-3}$, the Bondi mass
accretion rate is $\dot M_{B}=0.046 M_{\odot}$ yr$^{-1}$ for $\gamma=5/3$ 
(as adopted hereafter; \(\dot M_{B}\) would be 4.5 times larger if 
\(\gamma=1\); \citealt{2002apa..book.....F}). Taking into account the 
uncertainties on gas density, temperature and SMBH mass, and maximizing 
their effect on the computation of $\dot M_{B}$, the resulting range is 
$0.027<\dot M_{B} (M_{\odot}$ yr$^{-1})<0.072$. The accretion power is then 
$P_{B}=0.1\dot M_{B}c^2=2.6^{+1.5}_{-1.1}\times 10^{44}$ erg s$^{-1}$.

The nucleus of NGC 4649 is extremely radiatively inefficient, with
$L_{bol,nuc}<<P_{B}$ \citep[as already noticed][]{2005ApJ...624..155P}; our
estimate of the accretion rate allows to establish the accretion modalities. 
In fact $\dot m_{B}=\dot M_{B}/\dot M_{\rm Edd}=4.6\times 10^{-4}${, 
a very low value} well within the radiatively inefficient accretion (RIAF) 
regime, that 
can take place when $\dot m=\dot M/\dot M_{\rm Edd}<<0.01$ 
\citep{1995ApJ...452..710N}. RIAF models are the viscous rotating analog of 
the spherical Bondi accretion, with an efficiency for producing radiation of
$\epsilon \sim 10\dot m $. The expected $P_{RIAF}$ is then $\sim 10 \dot 
m_B \dot M_{B} c^2 =1.2^{+1.8}_{-0.8}\times 10^{43}$ erg s$^{-1}$, that is 
$\sim 3000$ times larger than $L_{bol,nuc}=4.2\times 10^{39}$ erg s$^{-1}$. 
Adopting instead a bolometric correction factor specific for the spectral 
energy distribution of a RIAF, i.e., $L_{0.5-8 {\rm keV}}\la 0.15 L_{bol,RIAF}$
\citep{1997ApJ...477..585M}, then $L_{bol,RIAF}\ga 4.8\times 10^{39}$ erg 
s$^{-1}$, which is $\la 2500$ times lower than the predicted $P_{RIAF}$. 
Reductions of the mass accretion rate with respect to the mass available at 
large radii (i.e., $\dot M_B$) on the way to the SMBH are possible, since 
RIAF solutions include cases of outflows or convective motions 
\citep{1999MNRAS.303L...1B}; another source of reduction is given by the
possibility that the gas has non-negligible angular momentum at the Bondi 
radius \citep{2003ApJ...592..767P,2011MNRAS.415.3721N}. The latter authors 
calculated the rate at which mass accretes onto a SMBH from rotating gas, 
for plausible RIAF solutions for galactic nuclei, and found that $\dot M\sim 
(0.3-1)\dot M_{B}$.  Indeed, the stellar component of NGC 4649 is known to
possess a significant rotation \citep{2003ApJ...596..903P}. If $\dot M=0.3 
M_{B}$, the predicted $P_{RIAF}$ goes down to $\sim 10^{42}$ erg 
s$^{-1}$, which is now larger than $L_{bol,RIAF}$ by a factor 
{between one hundred and six hundreds, considering the range for 
\(\dot M_{B}\).}

In conclusion, the accretion flow seems to emit less than predicted from the 
fuel observed to be available, even when allowing for a RIAF with angular 
momentum at the outer radius of the accretion flow (the various powers and luminosities considered here are summarized in Table \ref{table:powers}); possible solutions could 
be that the bolometric correction is larger, so that $L_{bol,nuc}$ becomes 
larger (this is not expected to fix entirely the problem, though), or that 
the accretion flow should be modeled differently (i.e., with an even lower 
radiative efficiency), or that outflows/convective motions are important. 
We note that a very low radiation efficiency was also found in the deep 
\(\sim 1\) Ms \textit{Chandra} observations of NGC 3115 
\citep{2014ApJ...780....9W,2014ApJ...782..103S} indicating for this source 
either a remarkably inefficient flow or a very low accretion rate 
suppressed by outflow, rotational support or stellar feedback.

However, the total (observed) power output from the black hole is the sum of 
the radiating (e.g., $L_{bol,nuc}$) and mechanical powers, where the latter 
by far dominates usually in the low luminosity nuclei of the local universe. 
Therefore, a further important constraint on how much mass must be accreting 
comes from $P_{cav}$, and is examined below.

\subsection{Jet power and total accretion output}\label{sec:jet_power}

Pointed VLA observations by \citet{2008MNRAS.383..923S} showed the radio 
properties at 1.4 GHz (see Fig. 3), and their connection with the cavities 
in the hot ISM (from a shallower 37 ksec \textit{Chandra} exposure).
\citeauthor{2008MNRAS.383..923S} derived a total flux density of $2.82\times 
10^{-2}$ Jy at 1.4 GHz; for a distance of 16 Mpc, this gives a radio power 
of $P_{1.4}=8.64\times 10^{27}$ erg s$^{-1}$ Hz$^{-1}$, that is 
$L_{1.4}=1.2\times 10^{37}$ erg s$^{-1}$. Thus NGC 4649 is a low-power radio 
source, one of the least radio emitting giant ellipticals with respect to 
those in the samples that produced the correlations between $\dot M_{B}$, 
$P_{cav}$, and radio luminosity mentioned above; it is useful then to 
establish whether extrapolation to lower luminosity sources is applicable.

Following \citet{2006MNRAS.372...21A}, \citet{2008MNRAS.383..923S} used the 
1.4 GHz image to determine the edges of the cavities, and then calculate the
kinetic energy of the jet, as the sum of their internal energy and the PdV 
work done to inflate them (as \(E=4PV\)). Since the majority of the energy 
carried off by these jets is mechanical, not radiative 
\citep[e.g.][]{2007MNRAS.381..589M}, the resulting energy \(E\), when 
divided by  an approximate age for the cavities, gives an estimate of (a 
lower limit to\footnote{Shocks and sound waves have also been found to 
contribute significantly to the power output of the central AGN 
\citep[e.g.][]{2006MNRAS.366..417F,2007ApJ...665.1057F,2011ApJ...726...86R}.}) the jet power $P_{cav}$. When considering both age estimates for the 
cavities, based on the buoyancy rise-time and sound-speed expansion 
timescales, $P_{cav}$ is in the range $(1-2)\times 10^{42}$ erg s$^{-1}$. 
This number is {$\sim 0.003-0.009$} of the accretion power $P_{B}$ derived
above. 

We calculated the cavity power from our deeper image adopting the same 
method of \citeauthor{2008MNRAS.383..923S}, but rather than using the radio 
contours as a guide we used instead the X-ray residual map to delineate the 
cavities (see Fig. \ref{fig:sign3}). These are approximated as ellipsoids 
centered {at} distance \(R\) from the SMBH and with semi-axes \(r_l\) along 
the radio jet direction (\(31\degree\) east of north) and \(r_w\) across it 
(the semi axis along the line of sight is taken equal to \(r_w\)). The 
regions shown in Fig. \ref{fig:sign3} have \(R=18.1''\), \(r_l=7.6''\), 
\(r_w=10.1''\) and \(R=14.1''\), \(r_l=7.5''\), \(r_w=5.2''\) for the N and 
S cavities, respectively, with a \(10\%\) uncertainty on these sizes. 
We evaluated an energetic content \(E_N=(1.9\pm0.8)\times{10}^{55}\mbox{ 
erg}\) and \(E_S=(0.9\pm 0.4)\times{10}^{55}\mbox{ erg}\) for the north and 
south cavity, respectively. The sound-speed expansion time has been 
evaluated as \(t_{cs}=r_l/c_s\), where \(c_s=\sqrt{\gamma k T/\mu m_p}\) 
\citep{2004ApJ...607..800B}, yielding \(t_{cs}=1.37 \pm 0.14 
\times{10}^6\mbox{ yr}\) and \(1.36 \pm 0.14 \times{10}^6\mbox{ yr}\) for 
the N and S cavity, respectively. Buoyancy rise times are evaluated as 
\(t_{buoy}=R/\varv_{buoy}\), where \(\varv_{buoy}\propto\sqrt{M(<R)\,r_l} 
/R\). Using \(M(<R)\) from Eq. \ref{equ:he} we obtain \(t_{buoy}=2.66 \pm 
0.41 \times{10}^6\mbox{ yr}\) and \(1.85 \pm 0.28 \times{10}^6\mbox{ yr}\) 
for the N and S cavity, respectively.

We then obtained for the two N and S cavities, for the sound speed expansion 
time $P_{cav,N}=(4.47 \pm 1.88) \times 10^{41}$ erg s$^{-1}$ and 
$P_{cav,S}=(1.99 \pm 0.84) \times 10^{41}$ erg s$^{-1}$, and for the 
buoyancy rise time $P_{cav,N}=(2.31 \pm 1.01) \times 10^{41}$ erg s$^{-1}$ 
and $P_{cav,S}=(1.46 \pm 0.65) \times 10^{41}$ erg s$^{-1}$. The two 
estimates are compatible between the errors, and give a total $P_{cav} 
\approx (3.8 \pm 1.2) \times 10^{41}$ erg s$^{-1}$ (for the buoyancy rise 
time), and $P_{cav}\approx (6.5 \pm 2.1) \times 10^{41}$ erg s$^{-1}$ (for 
the sound speed expansion time). Note that, as mentioned above, this may not 
be the whole of the mechanical power injected in the hot ISM; shocks are not 
seen here, though, possibly sound waves. Moreover, in addition to the larger 
cavities close to the ends of the jets, there are other smaller cavities 
seen in the X-ray gas (as well as other smaller radio structures, see Fig. 2 
of \citeauthor{2008MNRAS.383..923S}). The total of \(P_{cav}\) is then such 
that \(P_{cav}/P_B \gtrsim 0.0025\).

In the correlation found by \citet{2006MNRAS.372...21A} between $P_{cav}$ 
and $P_B$ for a sample of elliptical galaxies, on average $P_{cav}\approx 
0.2\,P_B$, a much larger fraction than found here. For NGC 4649, that has 
$P_B$ equal to the largest one in the \citeauthor{2006MNRAS.372...21A}'s 
sample, a larger $P_{cav}$ is predicted from our $P_B$ by their correlation: 
$P_{cav}=9.9\times 10^{43}$ erg s$^{-1}$ (with $P_{cav}=3.8\times 10^{43}$ 
erg s$^{-1}$ the lowest value allowed by all the uncertainties on $\dot M_B$ 
and on the correlation coefficients), about 100 times larger the jet power 
we estimate above. The \citeauthor{2006MNRAS.372...21A}'s relation implies 
that a significant fraction of $P_B$ emerges as kinetic energy, which 
requires that a significant fraction of the mass entering $r_{acc}$ flows 
all the way down to the SMBH, with little mass loss (in outflows) along the 
way. This seems not to be the case here, consistent with the results of the 
previous section, where a significant reduction of $\dot M$ with respect to 
$\dot M_B$ was needed. A similar result is obtained if one were to infer the 
accretion rate from the ``observed'' $L_{nuc,bol}=4.2\times 10^{39}$ erg 
s$^{-1}$. For the RIAF low radiative efficiency, one would derive $\dot 
M=8.5\times 10^{-4}M_{\odot}$ yr$^{-1}$, that is $\dot M\sim 0.02\,\dot 
M_B$. Such a value for $\dot M$ gives $P\approx 0.1\,\dot M c^2=4.9\times 
10^{42}$ erg s$^{-1}$, which can still account for the measured $P_{cav}$ 
(and is much closer to it, $P_{cav}/P\sim 0.1$). 
It seems so
that the same reduction in $\dot M$ required by the (far smaller) nuclear 
luminosity could explain also the (far larger) accretion output observed in 
mechanical form. In addition we note that in principle $P_{cav}$ is an average over the 
$\sim 1.4-2.7\times 10^6$ yr age of the cavities, while $L_{bol,nuc}$ is a 
measurement of the instantaneous accretion rate (since the  flow time from 
$r_{acc}$ to the SMBH is very small, of the order of $\sim 2\times{10}^5$ 
yr). Thus when linking $\dot M$ to $P_{cav}$ we assume that the accretion 
flow has been the same over the past \(\sim\) few \(10^6\) yr.
We note that 
\citet{2013MNRAS.432..530R} were not able to reproduce 
\citeauthor{2006MNRAS.372...21A}'s relation. This is mainly due to 
a different method for delineating the cavities adopted in the two papers, 
that is, radio maps for \citeauthor{2006MNRAS.372...21A} and X-ray maps for 
\citeauthor{2013MNRAS.432..530R} (the same method adopted in the present 
paper). As a result, the trend between cavity power and Bondi power 
appears significantly weakened in \citeauthor{2013MNRAS.432..530R} analysis 
(see their Figure 11) and with a larger scatter due to the different 
profiles adopted by these authors to extrapolate the gas density profiles 
to the accretion radius.

{Note that NGC 4649 does not display extended optical emission line regions and is depleted of molecular and atomic gas, as shown by \citet{2002AJ....124..788Y}, that put an upper limit of \({6.8}\times{10}^7\,M_{\astrosun}\) to the \(\mbox{H}_2\) mass content of this galaxy. If we assume that the AGN mechanical power we evaluate in the cavities is generated by accretion at \(\sim 0.1 \dot{M} c^2\), the total accreted mass to power the jet would be \(\sim 300\,M_{\astrosun}\), well below the upper limit evaluated by \citeauthor{2002AJ....124..788Y}. However, in NGC 4649 there is (as in general for hot gas rich galaxies) a large hot gas reservoir \(\sim 3\times{10}^9\,M_{\astrosun}\), much larger than the cold gas upper limit. Then, on a secular scale it is more likely that accretion is taking and took place thanks to the hot gas. Even if the primary source of accretion is likely to be the hot gas, however, the low accretion rate we evaluate does not exclude molecular gas as a source of fuel for the SMBH, so cold accretion of molecular and atomic gas represents a possible alternate (or additional) scenario to Bondi accretion.}

\subsection{Jet power, radio and nuclear luminosity}

\citet{2010ApJ...720.1066C} re-examined the scaling relationship between jet 
power, $P_{jet}$, and synchrotron luminosity at 1.4 GHz and 200-400 MHz, 
incorporating measurements for 21 giant elliptical galaxies, thus expanding 
the sample of \citet{2008ApJ...686..859B}, dominated by bubbles in clusters, 
to lower radio power sources. The $P_{jet}$ was estimated as $P_{cav}$, the 
power required to inflate the cavities. The 
\citeauthor{2010ApJ...720.1066C}'s relationship predicts $P_{cav}=5.2\times 
10^{41}$ erg s$^{-1}$ from $L_{1.4}$ of NGC 4649, and between $8.7\times 
10^{40}$ and $3.2\times 10^{42}$ erg s$^{-1}$ considering the large scatter 
around the correlation\footnote{The $P_{cav} - P_{1.4}$ relation shows a 
large scatter, possibly due to the fact that the radio emission from lobes 
depends on their composition, and processes such as gas entrainment, shocks 
and aging. In the case of NGC 4649 we do not see evidence of shocks. Shurkin 
et al. investigated the particle content of the cavities by determining k/f, 
where k is the ratio of the total particle energy contained in the cavity to 
the energy accounted for by electrons emitting synchrotron radiation in the 
range of 10 MHz to 10 GHz, and f is the volume filling factor of the 
relativistic plasma in the cavity. In general, k/f is in the range of \(\sim 
1\) to \(\sim 1000\) for cavities that are active, while the values for NGC 
4649 are very large ($\sim 10^4$). This could be explained by entrained 
particles, or electrons that have aged, which is likely given the small 
radio flux; NGC 4649 could soon become a ghost cavity system.} of 0.78 dex. 
This range of $P_{cav}$ includes the direct estimate from Sect. 
\ref{sec:jet_power} (see Figure \ref{fig:cavagnolo}, left panel). Similar 
results are obtained for the \citet{2011ApJ...735...11O} $P_{cav} - L_{1.4}$ 
relation. Thus, NGC 4649 seem to host cavities similar in nature to those of 
other giant ellipticals, at least for their radio properties, i.e., the 
relation between radiative and mechanical energy of the jets.

The nuclear 2-10 keV luminosity of NGC 4649 also fits on the correlation 
between nuclear 2-10 keV luminosity and $P_{cav}$, derived for 27 detected 
nuclei of central dominant galaxies (\citealt{2013MNRAS.432..530R}; see also 
\citealt{2007MNRAS.381..589M}). The majority of the sample is in a RIAF 
mode, and the mechanical cavity power dominates the radiative output. NGC 
4649 sits at the lowest values of the correlation, with \({L_{2-10\,{\rm 
keV}}} = 4.2\times{10}^{38}\mbox{ erg} \mbox{ s}^{-1}\), from which we can 
predict $P_{cav}\sim 4\times 10^{41}$ erg s$^{-1}$ (see Figure 
\ref{fig:cavagnolo}, right panel). Even though also in this correlation the 
scatter is large, it was taken as evidence that the radiative efficiency of 
the X-ray nucleus increases with increasing $P_{cav}$, until the quasar 
regime is reached (where the nuclear luminosity becomes comparable to
$P_{cav}$).

\section{Summary and Conclusions}\label{sec:summary}

We investigated the presence of AGN feedback in the ISM of the giant 
elliptical NGC 4649 by using a total of 280 ks \textit{Chandra} observations
This source has been studied several times in different wavelength, and in 
particular in the X-rays 
\citep[e.g.][]{2010MNRAS.404.1165C,2010MNRAS.409.1362D,2012ApJ...757..121L}.
\citet{2008MNRAS.383..923S} and \citet{2008ApJ...683..161H}, making use of 
\textit{Chandra} observations, studied the properties of the ISM in NGC 
4649 using the unsharp-mask technique looking for morphological 
disturbances pointing to deviations from the hydrostatic equilibrium 
condition suggested by the generally relaxed X-ray morphology. 
Interestingly, while the former authors using a shallow \(\sim 37\mbox{ 
ksec}\) observation found evidences of structures and cavities in the ISM 
that they interpreted as connected with the central, faint radio source, 
the latter authors using deeper \(\sim 81 \mbox{ ksec}\) data did not found 
any evidence of such disturbances. A subsequent analysis by 
\citet{2010MNRAS.404..180D} showed disturbances and cavities in the ISM as 
residuals of the X-ray surface brightness from a spherical \(\beta\) model, 
connected with the radio emission.

\begin{figure}
\centering
\includegraphics[scale=0.40]{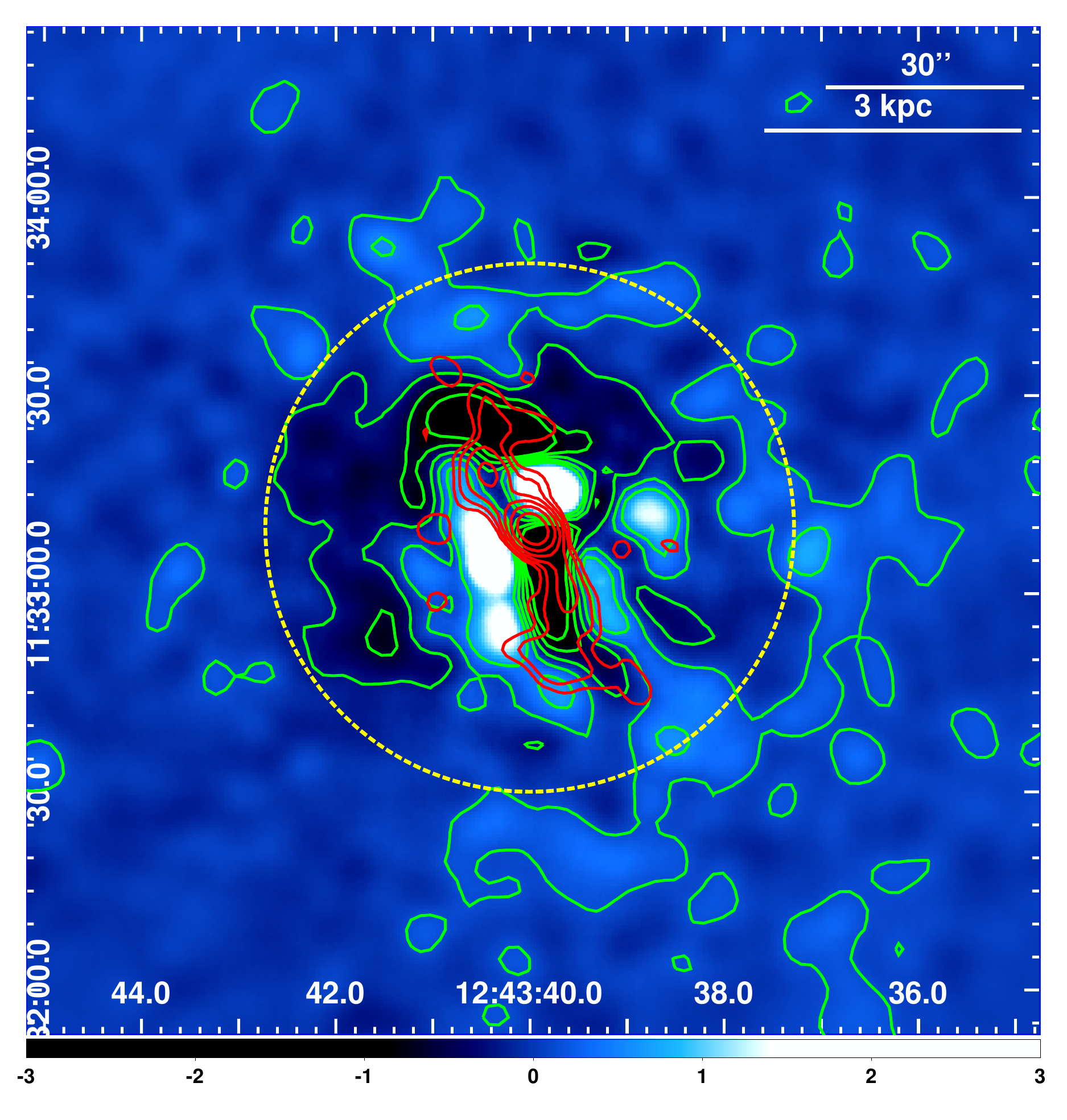}
\includegraphics[scale=0.40]{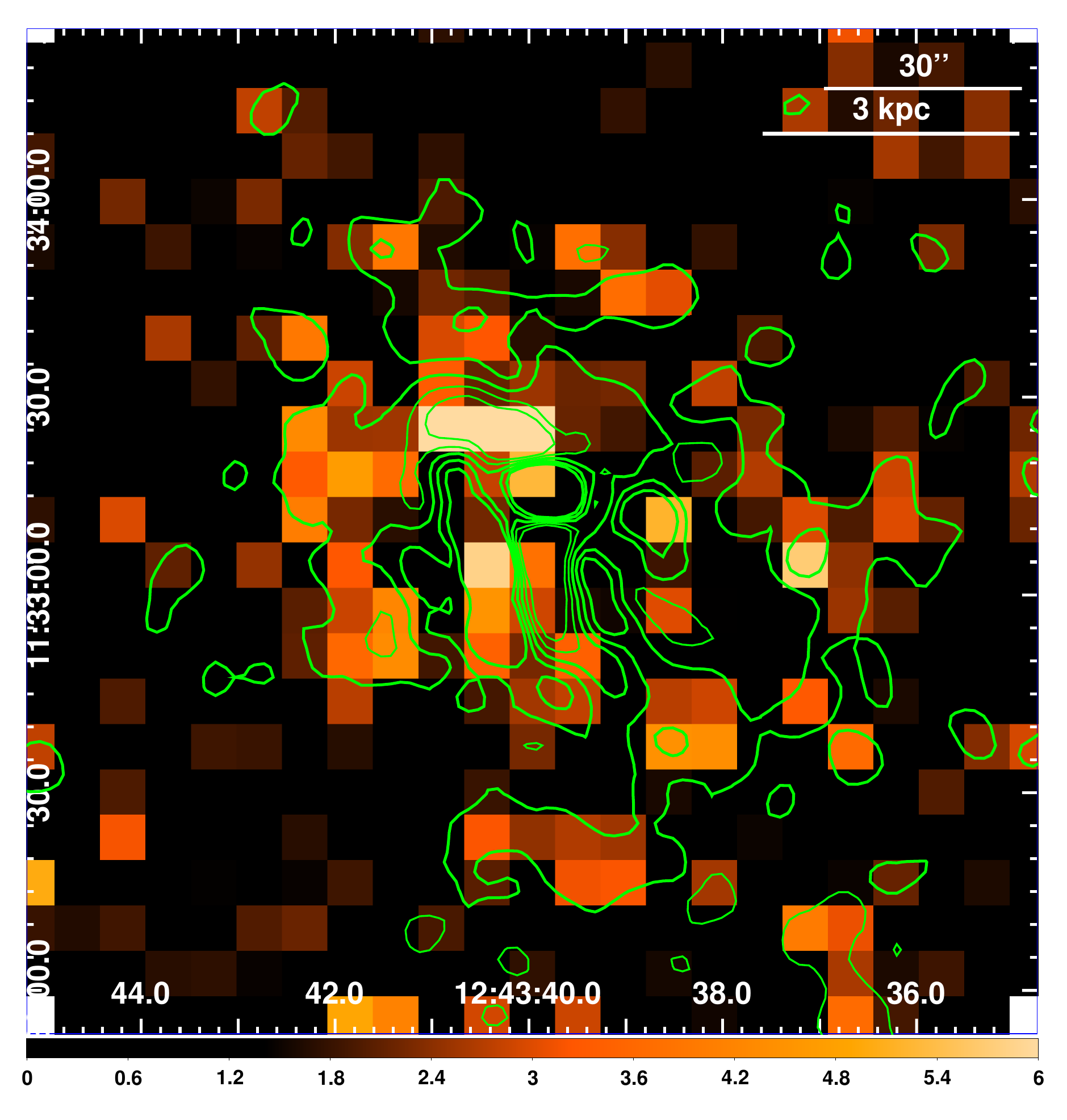}
\caption{Same of Figure \ref{fig:sign3}, but on a larger scale \(\sim 
5\mbox{ kpc}\). (left panel) Distribution of residuals relative to an 
elliptical \(\beta\) model, with a 10X10 FWHM gaussian smoothing, with 
contours shown in green and VLA emission contours are shown in red. The 
dashed yellow circle indicates the \(\sim 40''\) radius where we see 
significant residuals in the one-dimensional surface brightness fit (see 
Figure \ref{rprofile}). (right panel) Residual S/N map, evaluated as the 
ratio between the residuals and the X-ray counts error, binned to a 
pixel size 14 times the size of the native ACIS-S pixel. Green contours are 
the same as in left panel.}\label{fig:sign10}
\end{figure}

\begin{figure}
\centering
\includegraphics[scale=0.43]{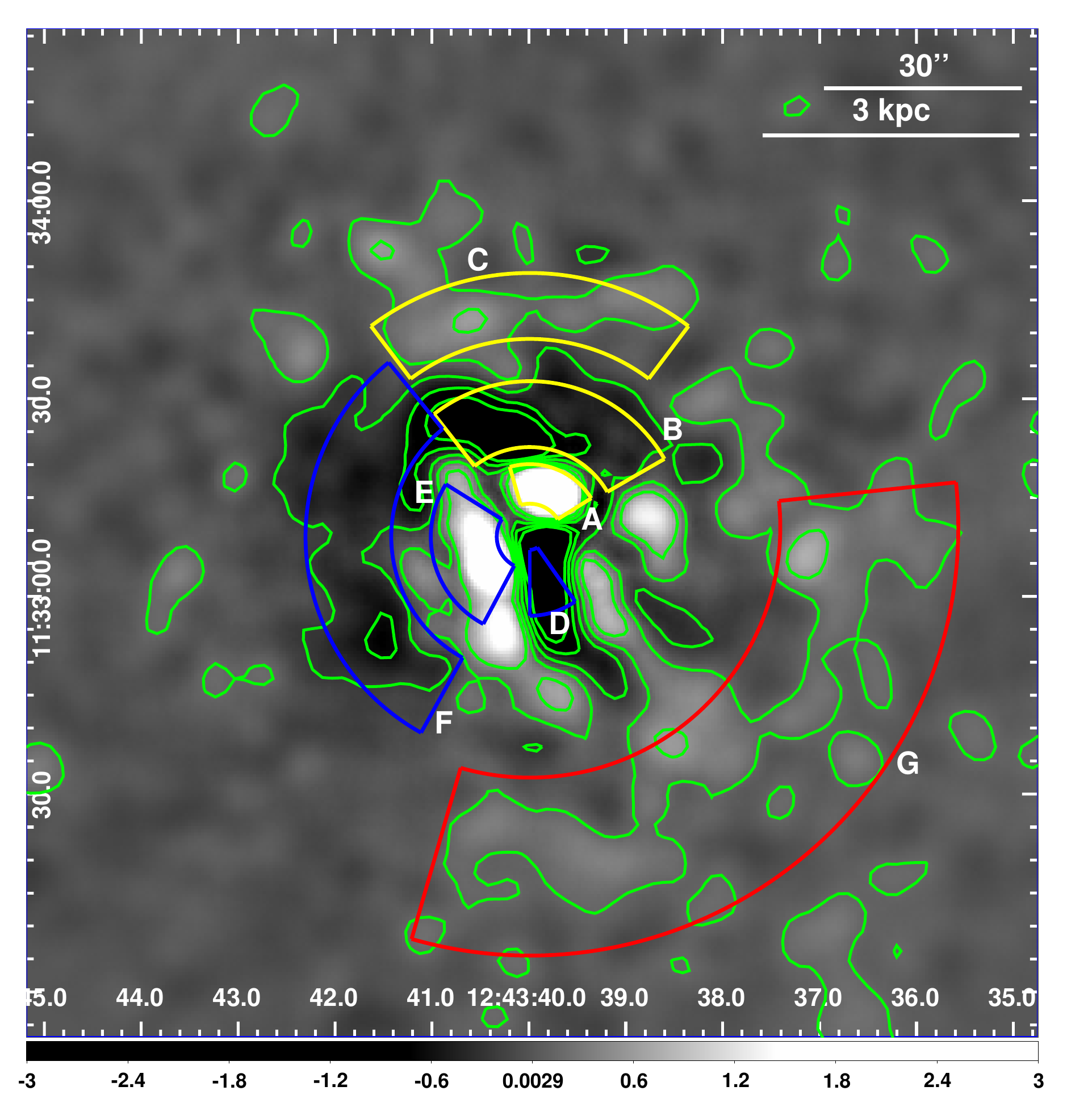}
\caption{Same as left panel of Figure \ref{fig:sign10} with superimposed 
the spectral extraction regions discussed in Section 
\ref{sec:spectra}.}\label{fig:outer_regions}
\end{figure}

\begin{figure}
\centering
\includegraphics[scale=0.6]{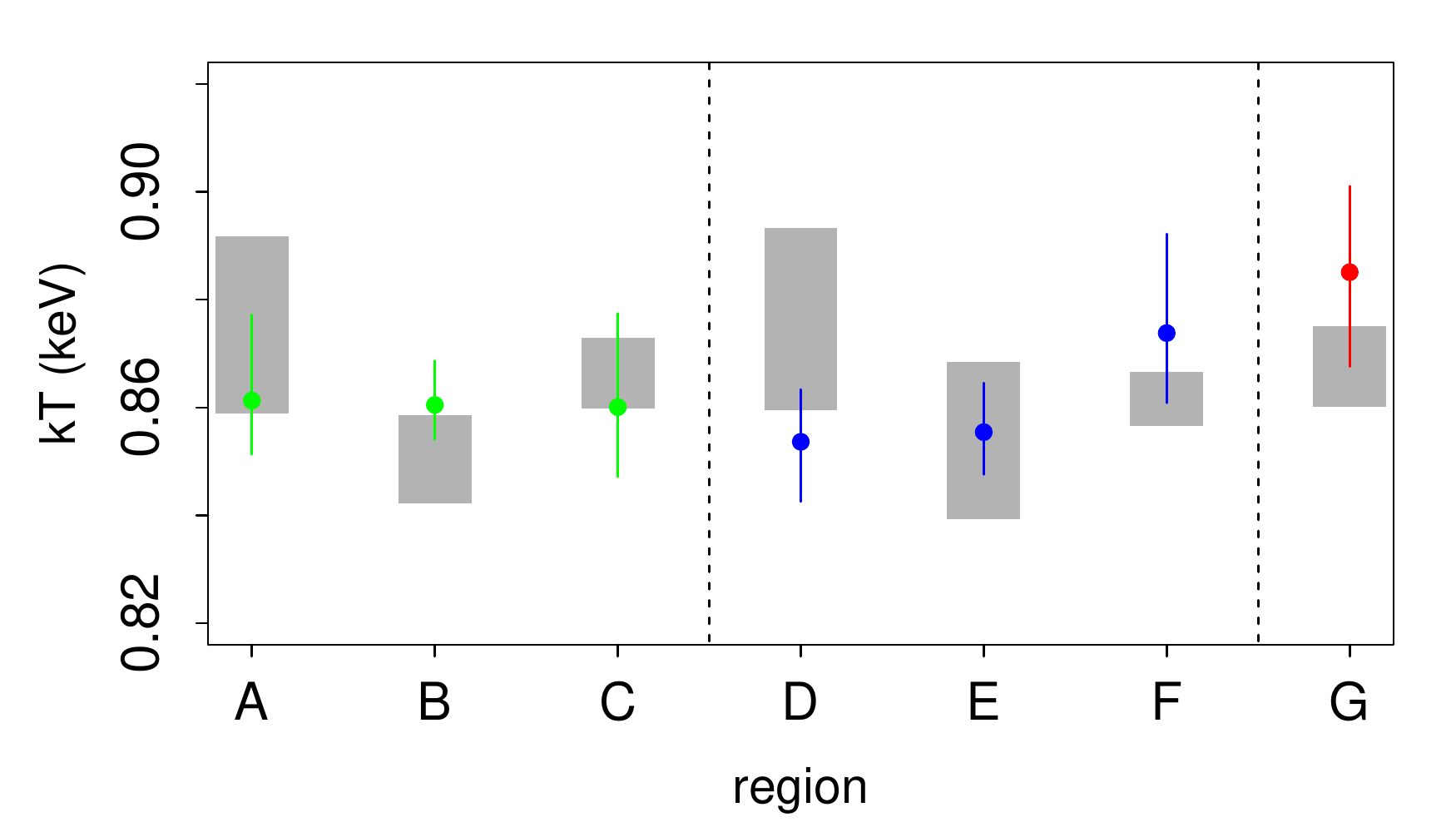}
\includegraphics[scale=0.6]{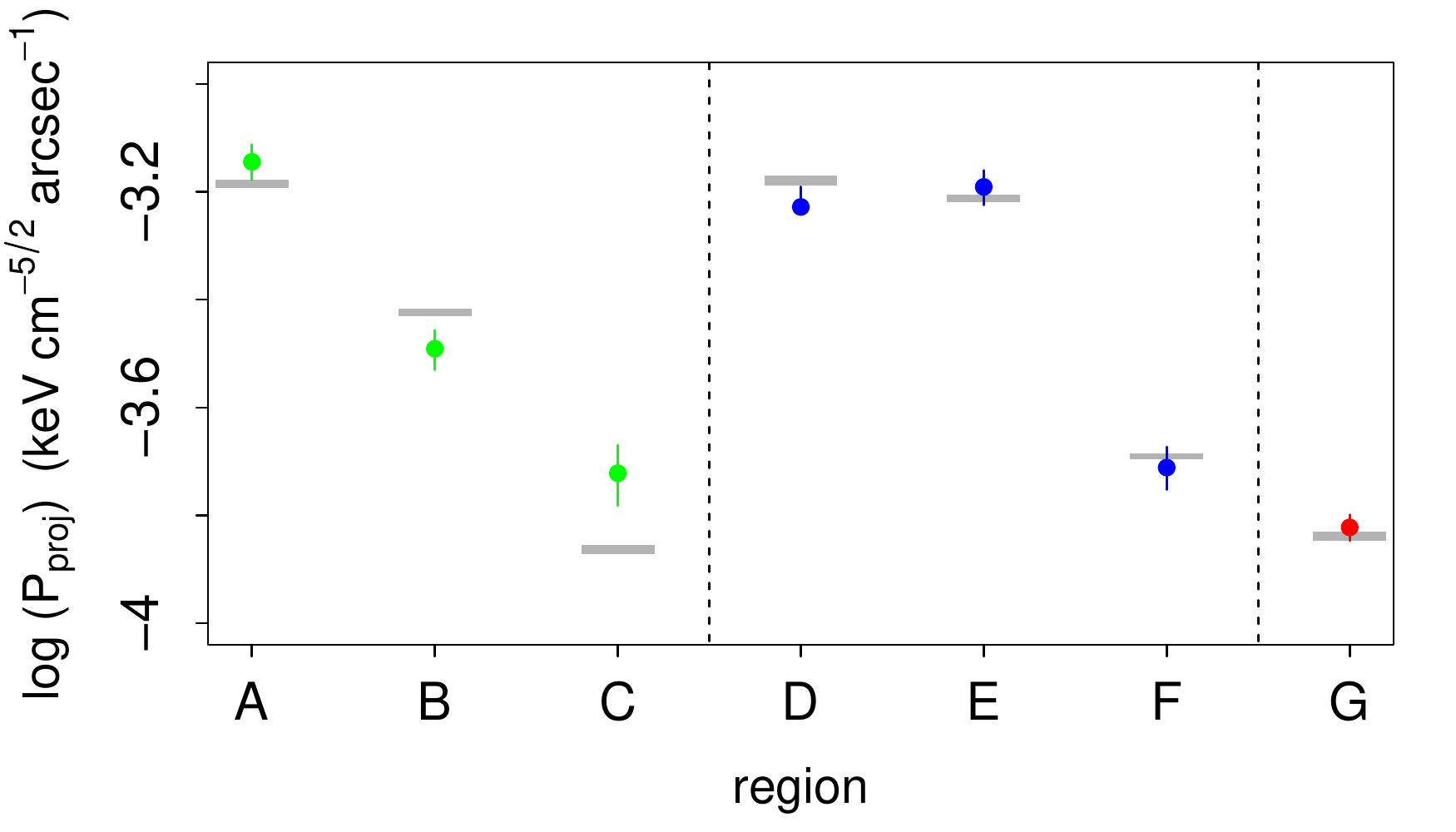}
\caption{Spectral parameters from Table \ref{table:fit_results}. The 
parameters are shown with colored points (according to the region colors 
shown in Figure \ref{fig:outer_regions}) with errors reported in Table 
\ref{table:fit_results}, while in grey we show the average values of the 
same parameters evaluated in concentric annuli at the same radius (see 
Section \ref{sec:spectra}).}\label{fig:spectral_results}
\end{figure}

\begin{figure}
\centering
\includegraphics[scale=0.43]{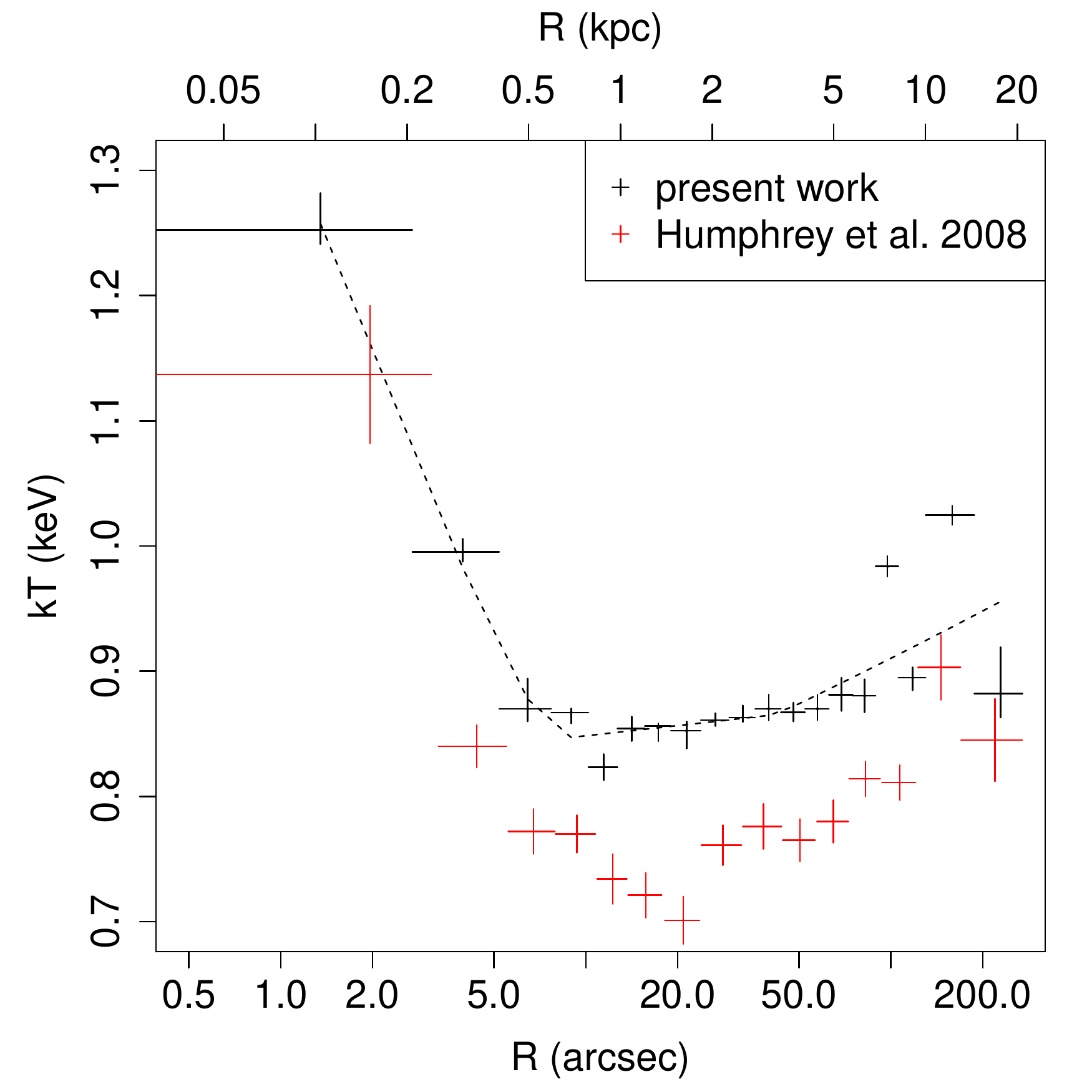}
\includegraphics[scale=0.43]{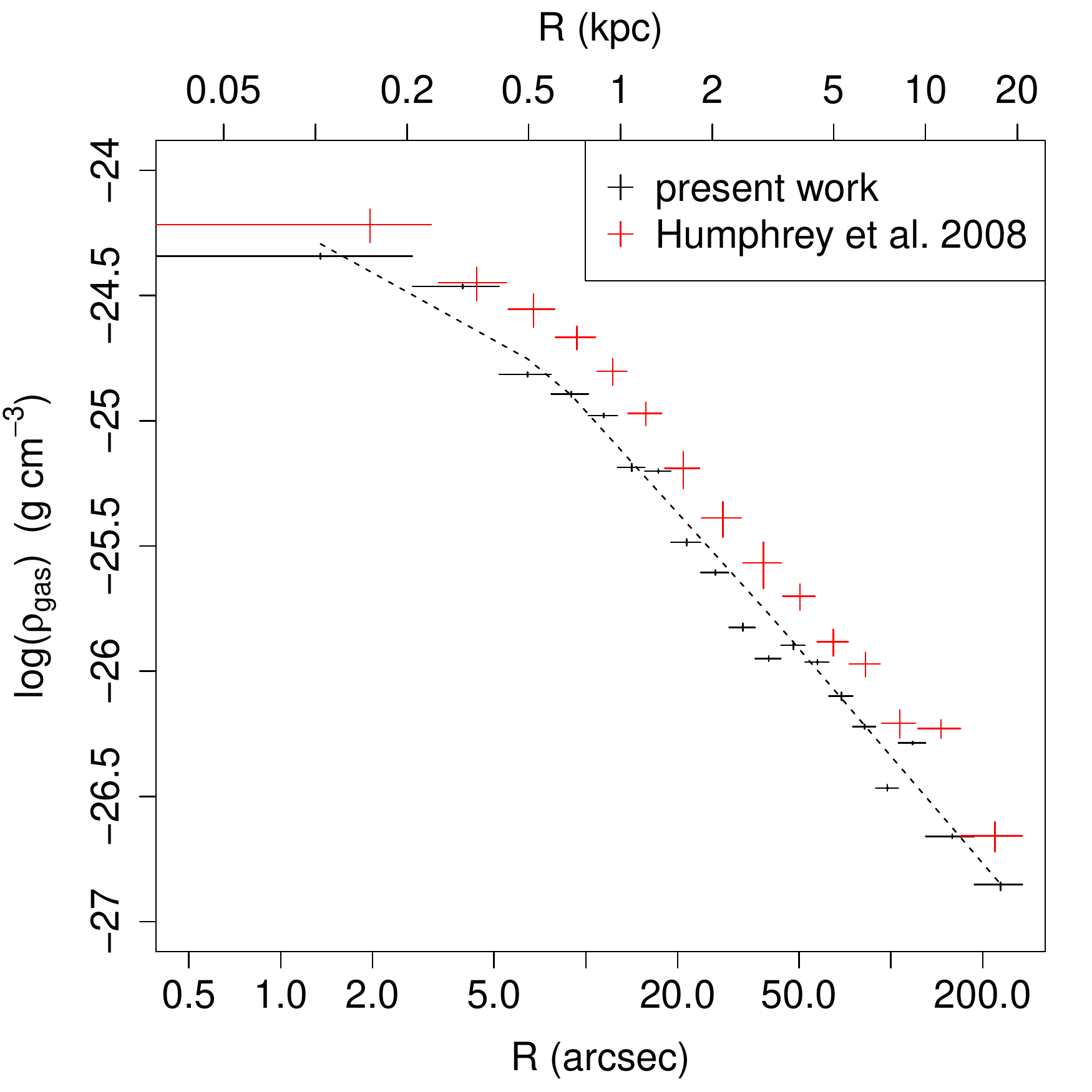}
\caption{Gas temperature (left panel) and density (right panel) derived 
from the concentric annuli fit described in Section \ref{sec:spectra}. 
Black crosses represent the profiles obtained in this work (errors bar 
represent \(1\)-\(\sigma\) confidence level), while for comparison we show 
with red crosses the profiles obtained by \citep{2008ApJ...683..161H} with 
shallower data (see Section \ref{sec:spectra}).}\label{fig:profiles}
\end{figure}

\begin{figure}
\centering
\includegraphics[scale=0.43]{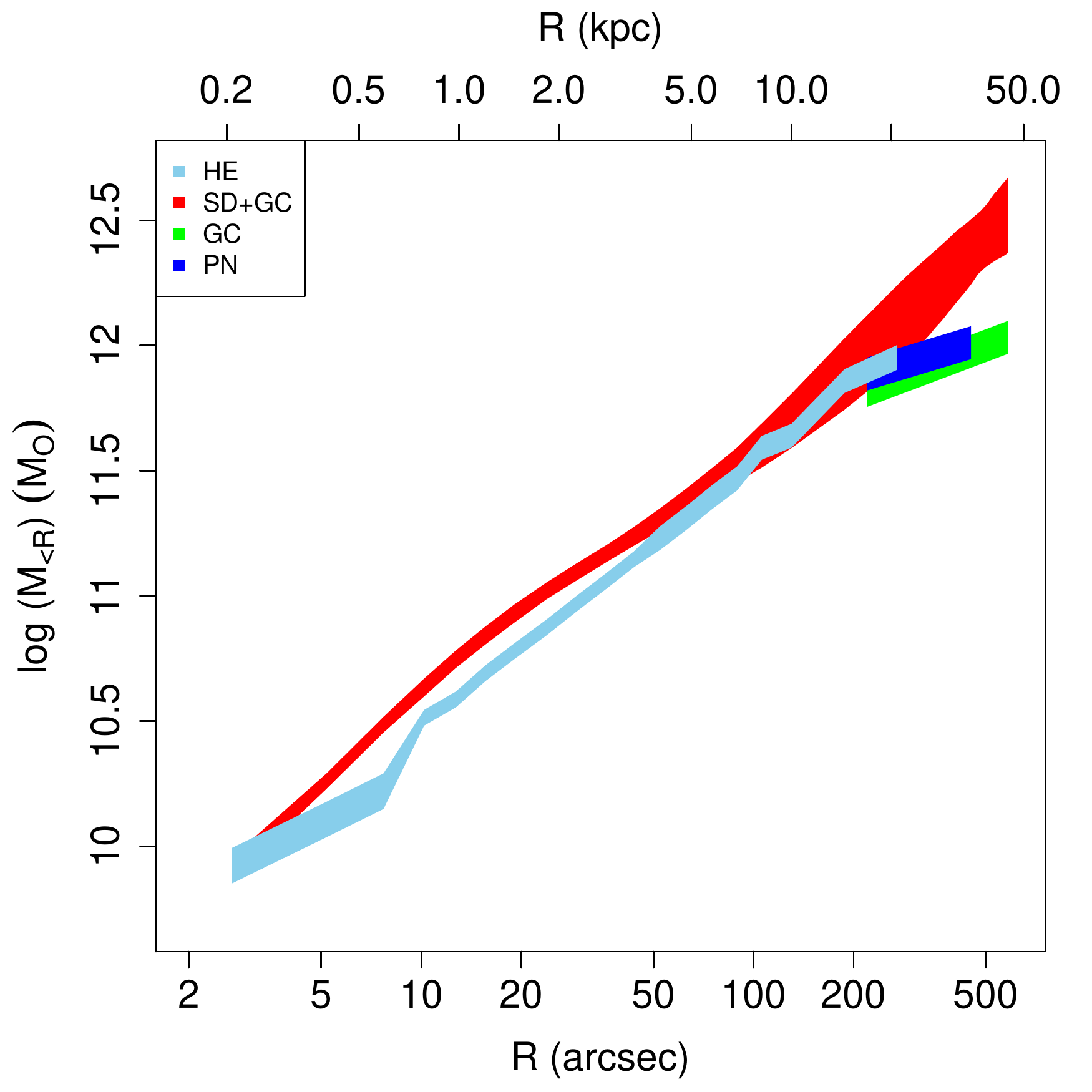}
\includegraphics[scale=0.43]{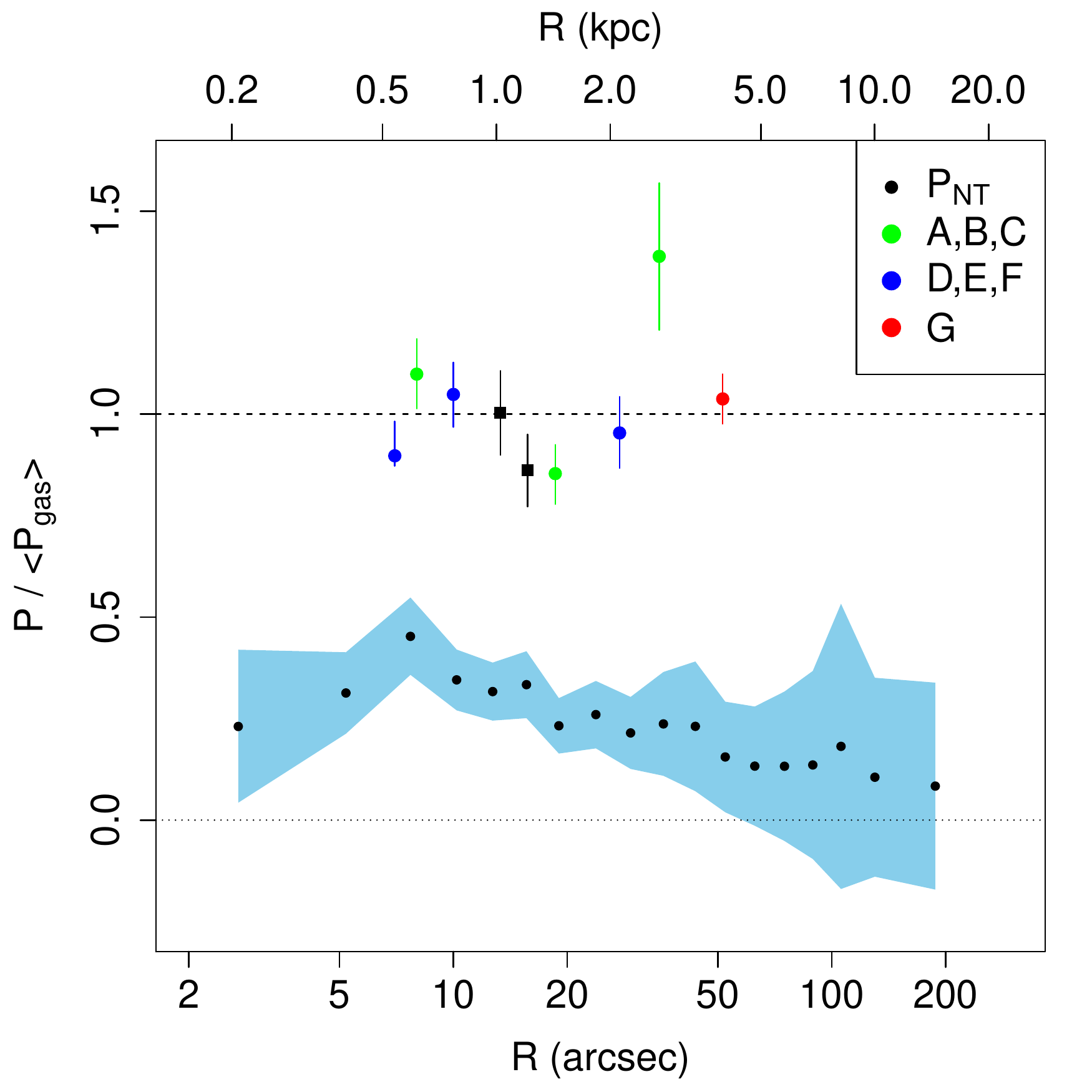}
\caption{(left panel) Mass profiles obtained from the hydrostatic 
equilibrium Eq. \ref{equ:he} (light blue strip) and from the stellar and GC 
kinematics by \citet[][red strip]{2010ApJ...711..484S}. Note the 
significant deviation between 0.5 and 3 kpc. The mass profiles obtained by 
\citet{2012ApJ...748....2D} using PNe and GC kinematics are shown with a 
blue and green strip, respectively. (right panel) Ratio of the non-thermal 
pressure component derived from Eq. \ref{equ:ntp} to the average gas 
pressure (black circles, with the light blue strip representing the 
uncertainty). For comparison, we also present with colored points the ratio 
of the gas pressure in the regions shown in Figure \ref{fig:outer_regions}, 
while the black squares represent the minimum radio pressure evaluated in 
Sect. \ref{sec:ntpress}.}\label{fig:masses}
\end{figure}

\begin{figure}
\centering
\includegraphics[scale=0.43]{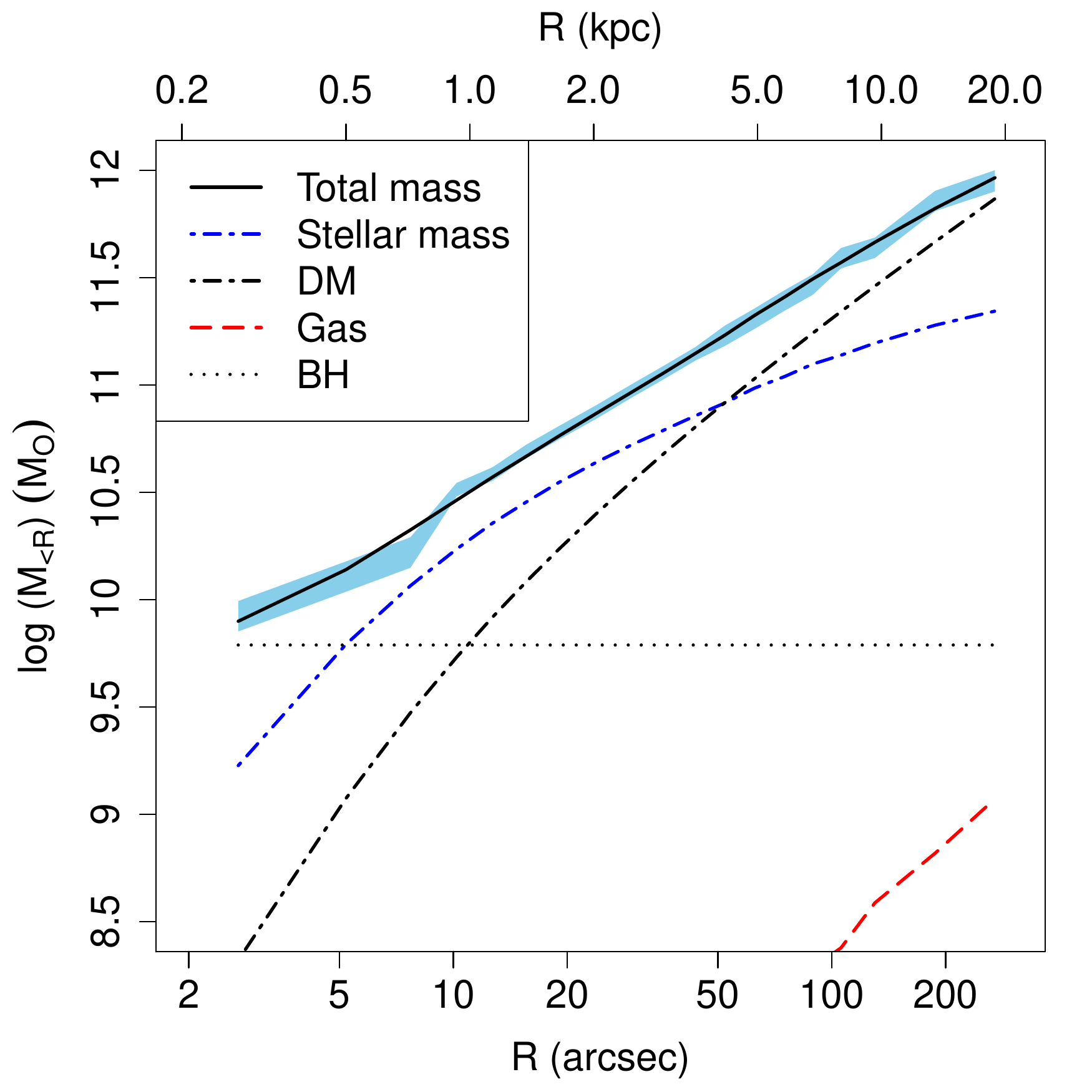}
\caption{Fit of the mass profile derived with Eq. \ref{equ:he} (blue strip) 
with the four standard galaxy mass components.}\label{fig:mass_fit}
\end{figure}

\begin{figure}
\centering
\includegraphics[scale=0.43]{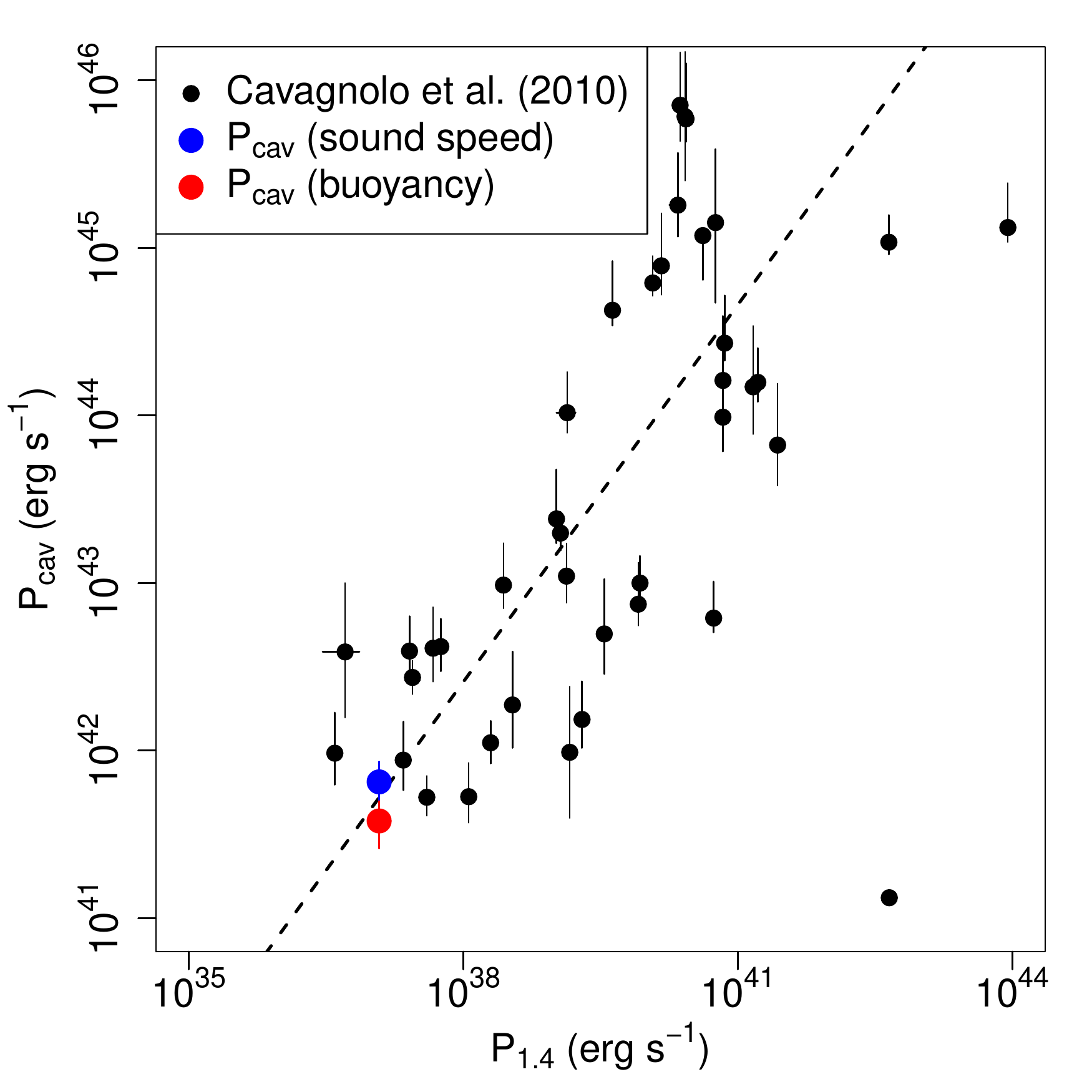}
\includegraphics[scale=0.43]{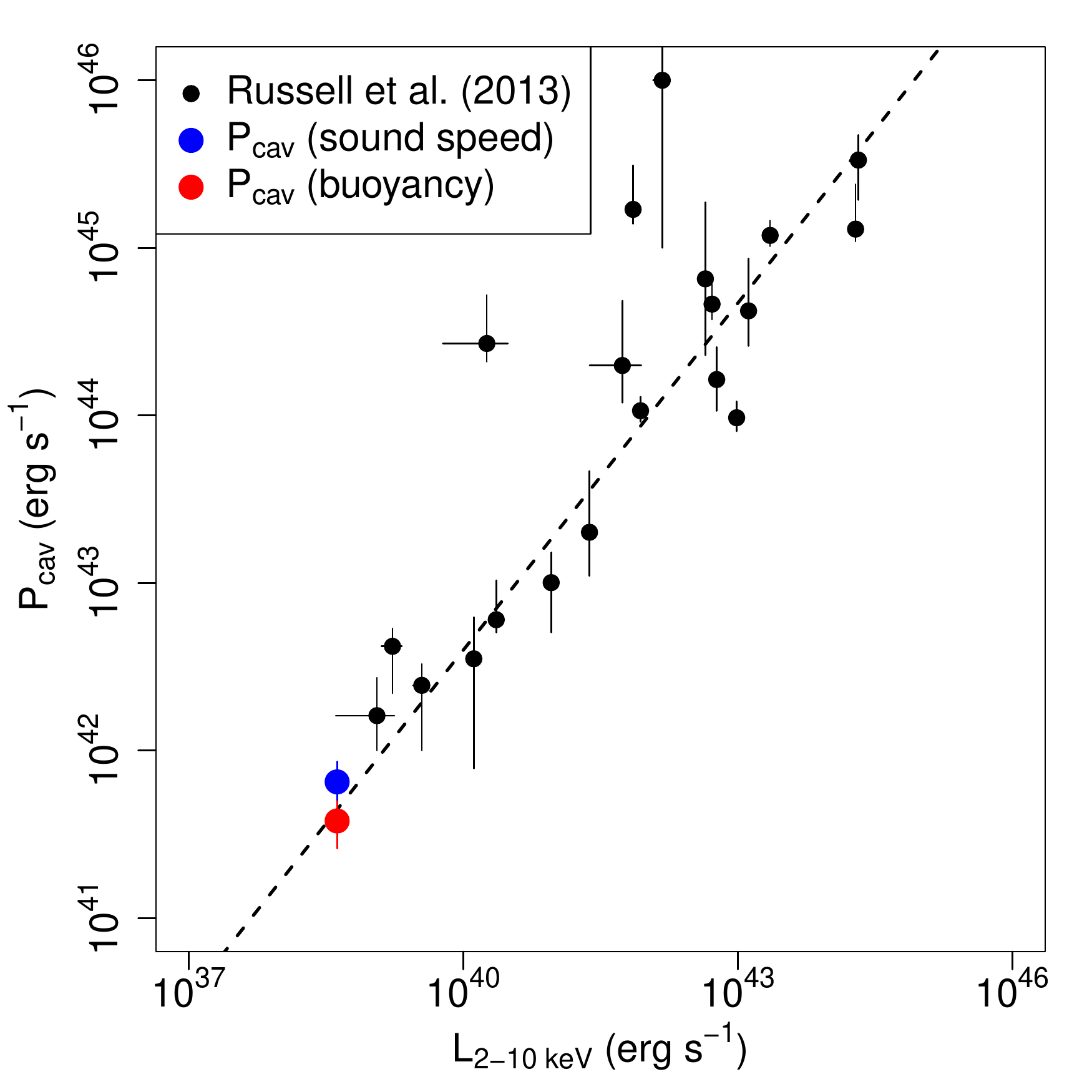}
\caption{(left panel) Cavity power vs. radio power in the sample of giant 
elliptical galaxies from \citet{2010ApJ...720.1066C} (black circles) and in 
NGC 4649 evaluated from the sound speed expansion time (blue circle) and 
from the buoyancy rise time (red circle). The dashed line represents the 
best fit relation from \citeauthor{2010ApJ...720.1066C}. (right panel) 
Cavity power vs. nuclear X-ray luminosity in the sample of bright galaxy 
cluster from \citet{2013MNRAS.432..530R} (black circles) and in NGC 4649 
(same as left panel). The dashed line represents the best fit relation from 
\citeauthor{2013MNRAS.432..530R}.}\label{fig:cavagnolo}
\end{figure}

Using much deeper \textit{Chandra} data with a total exposure \(\sim 
280\mbox{ ksec}\), we used the latter approach to investigate the 
morphological distribution of the ISM in NGC 4649. We studied the deviation 
of the X-ray surface brightness from an elliptical \(\beta\) model, which 
is expected to describe the hot gas distribution in relaxed galaxies. The 
residuals of this fitting procedure, presented in Figure \ref{fig:sign10}, 
show significant cavities, ripples and ring like structures on the inner 
\(0.5\div 3\mbox{ kpc}\) scale.
This is at variance with the \(\lesssim 2\sigma\) significance of the 
cavities reported by \citet{2008MNRAS.383..923S} as evaluated by 
\citet{2008ApJ...683..161H}. The deeper data considered here revealed these 
structures with high significance; moreover the structures appear to be 
morphologically related with the central radio emission, with cavities 
lying in correspondence with the extended radio lobes and regions of 
enhanced emission situated on the side of them and, on larger scale, taking 
the form of ring like ripples which seems reminiscent of the structures 
observed in NGC 1275 \citep{2006MNRAS.366..417F}. In common with this 
source, we found no significant temperature variations in correspondence 
with higher pressure regions. So, if radio ejecta driven shocks are 
responsible for the observed ISM morphology, the observed structures may be 
isothermal waves whose energy is dissipated by viscosity, with thermal 
conduction and sound waves effectively distributing the energy from the 
radio source. Evidences of deviations from the hydrostatic equilibrium are 
also provided by the mass profiles presented in Figure \ref{fig:masses} 
(left panel). A significant non-thermal pressure is found on the same scale 
of the residual structures, where it reaches \(\sim 30\%\) of the observed 
gas pressure. In addition, the excess gas pressure and non-thermal pressure profiles appear to be strongly correlated, indicating the radio ejecta as the likely origin for this additional pressure component. 

At smaller scales, similarly to a few other early type galaxies harboring 
low power radio sources, NGC 4649 shows increased temperatures in the inner 
\(0.5\mbox{ kpc}\) region. The nucleus of NGC 4649 appears to be extremely 
sub-Eddington, with the accretion flow emitting less than predicted from 
the fuel observed to be available, even when allowing for a RIAF with 
angular momentum at the outer radius of the accretion flow. Also the jet 
power evaluated from the observed X-ray cavities appears to be much smaller 
than that predicted for elliptical galaxies from the Bondi accretion power 
\(P_B\). If the mass accretion rate accounting for the observed nuclear 
X-ray luminosity is adopted - which requires, in addition to a low 
radiative efficiency, a significant reduction of the accretion rate with 
respect to the Bondi value, due, e.g., to outflows/convective motions - 
then the corresponding accretion power \(P\) is \(\sim 10\) times larger 
than the observed kinetic {power}. When comparing the 
jet power to radio and nuclear X-ray luminosity, on the other hand, the 
observed cavities show similar behavior to those of other giant elliptical 
galaxies.

\acknowledgments
{We acknowledge useful comments and suggestions by our anonymous referee.}
This work was partially supported by NASA contract NAS8-03060 (CXC), and NASA Chandra grant G01-12110X.
DWK acknowledges partial support from Smithsonian Institute 2014 Competitive Grants Program for Science.
SP acknowledges financial support from MIUR grant PRIN 2010-2011, project 
`The Chemical and Dynamical Evolution of the Milky Way and Local Group 
Galaxies', prot. 2010LY5N2T.
FC acknowledges financial support by the NASA contract 11-ADAP11-0218.
This research has made use of software provided by the Chandra X-ray Center (CXC) in the application packages CIAO, ChIPS, and Sherpa.

\newpage

\begin{table}
\caption{Observations properties.}\label{table:obs}
\begin{center}
\begin{tabular}{rccc}
\hline
\hline
Obs ID & Net Exposure (ksec) & Date    & PI Name \\
\hline
785    & 23.9            & 2000-04 & Sarazin \\
8182   & 47.8            & 2007-01 & Humphrey \\
8507   & 17.3            & 2007-02 & Humphrey \\
12976  & 98.1            & 2011-02 & Fabbiano \\
12975  & 79.8            & 2011-08 & Fabbiano \\
14328  & 13.4            & 2011-08 & Fabbiano \\
\hline
\hline
\end{tabular}
\end{center}
\end{table}

\begin{table}
\caption{Properties of the spectra extracted from the regions discussed in 
Section \ref{sec:spectra}. For each region we indicate its name, the net 
\(0.3-8\mbox{ keV}\) counts, the residual significance in the region after 
subtracting the \(\beta\) model, and best fit reduced \(\chi^2\), gas 
temperature and projected pressure with \(1\)-\(\sigma\) confidence 
errors.}\label{table:fit_results}
\begin{center}
\begin{tabular}{crcccc}
\hline
\hline
Region & Net Counts (error) & Residuals sign. & \(\chi^2\) (d.o.f.) & \(kT\) & 
\(P_{proj}\) \\
 & & & & \(\mbox{ (keV)}\) & \(({10}^{-4}\mbox{ keV cm}^{-5/2}\mbox{ 
 arcsec}^{-1})\) \\
\hline
A & 7659 (88) & 9 & 1.06 (90) & \({0.86}_{-0.01}^{+0.02}\) & 
\({7.17}_{-0.55}^{+0.55}\)\\
B & 8454 (93) & 11 & 1.03(99) & \({0.86}_{-0.01}^{+0.01}\) & 
\({3.22}_{-0.28}^{+0.27}\) \\
C & 3865 (65) & 6 & 1.00 (79) & \({0.86}_{-0.01}^{+0.02}\) & 
\({1.90}_{-0.25}^{+0.24}\) \\
D & 5840 (77) & 6 & 1.08 (76) & \({0.85}_{-0.01}^{+0.01}\) & 
\({5.92}_{-0.13}^{+0.54}\) \\
E & 14565 (121) & 7 & 1.08 (111) & \({0.86}_{-0.01}^{+0.01}\) & 
\({6.44}_{-0.48}^{+0.48}\) \\
F & 8125 (93) & 11 & 0.96 (102) & \({0.87}_{-0.02}^{+0.01}\) & 
\({1.94}_{-0.18}^{+0.18}\) \\
G & 11657 (117) & 10 & 1.07 (161) & \({0.89}_{-0.02}^{+0.02}\) & 
\({1.51}_{-0.08}^{+0.08}\) \\
\hline
\hline
\end{tabular}
\end{center}
\end{table}

\begin{table}
\caption{Power estimates evaluated in Sects. \ref{sec:bondi_power} and \ref{sec:jet_power}.}\label{table:powers}
\begin{center}
\begin{tabular}{lccc}
\hline
\hline
Name & Symbol & Value & Note \\
 & & \((\mbox{erg}\mbox{ s}^{-1})\) & \\
\hline
Bondi accretion power & \(P_B\) & \(2.6\times{10}^{44}\) & \(P_{B}=0.1\dot M_{B}c^2\) \\
Bolometric luminosity & \(L_{bol,nuc}\) & \(4.2\times{10}^{39}\) & \(L_{bol}/L_{2-10\, {\rm keV}} = 10\) \\
RIAF luminosity & \(P_{RIAF}\) & \(1.2\times{10}^{43}\) & \(P_{RIAF} = 10 \dot m_B \dot M_{B} c^2\) \\
RIAF luminosity + ang. mom. & - & \(1.2\times{10}^{42}\) & \(\dot M=0.3 M_{B}\) \\
RIAF bolometric luminosity & \(L_{bol,RIAF}\) & \(4.8\times{10}^{39}\) & 
\(L_{bol,RIAF}/L_{0.5-8\, {\rm keV}}\gtrsim 7\) \\
Cavity power & \(P_{cav}\) & \(3.8\times{10}^{41}\) & buoyancy rise  \\
'' & \(P_{cav}\) & \(6.5\times{10}^{41}\) & sound speed expansion \\
\hline
\hline
\end{tabular}
\end{center}
\end{table}

\end{document}